\documentclass[twocolumn,twocolappendix,superscriptaddress]{aastex631mod}

\usepackage{amsmath,amssymb}
\usepackage{fontawesome}
\usepackage{multirow}

\newcommand\CHIMERA{\texttt{CHIMERA}}
\newcommand{\OFour}{\textit{O4-like}}
\newcommand{\OFive}{\textit{O5-like}}
\newcommand{\zp}{\ensuremath{z_{\rm phot}}}
\newcommand{\zs}{\ensuremath{z_{\rm spec}}}
\newcommand{\Spectral}{Spectral}
\newcommand{\Fullzp}{Full (\zp)}
\newcommand{\Fullzs}{Full (\zs)}
\newcommand{\codestring}[1]{\textcolor{red!80!black}{\texttt{\textquotesingle #1\textquotesingle}}}

\newcommand\given[1][]{\,#1\vert\,}
\newcommand{\vlambda}{\boldsymbol{\lambda}}
\newcommand{\vtheta}{\boldsymbol{\theta}}
\newcommand{\vdataGW}{\boldsymbol{d}^{\rm GW}}
\newcommand{\vdataEM}{\boldsymbol{d}^{\rm EM}}

\newcommand{\msun}{\ensuremath{\mathrm{M}_\odot}}
\newcommand{\kmsMpc}{\ensuremath{\rm km\,s^{-1}\,Mpc^{-1}}}

\newcolumntype{X}[1]{>{\centering\let\newline\\\arraybackslash\hspace{0pt}}m{#1}}

\received{December 18, 2023}
\accepted{January 20, 2024}

\begin{document}

\title{\large Cosmology and Astrophysics with Standard Sirens and Galaxy Catalogs in View of Future Gravitational Wave Observations}

\author[0000-0002-2889-8997]{Nicola Borghi}
\affiliation{Dipartimento di Fisica e Astronomia ``Augusto Righi''--Universit\`{a} di Bologna, via Piero Gobetti 93/2, I-40129 Bologna, Italy}
\affiliation{INAF - Osservatorio di Astrofisica e Scienza dello Spazio di Bologna, via Piero Gobetti 93/3, I-40129 Bologna, Italy}

\author[0000-0002-0675-508X]{Michele Mancarella}
\affiliation{Dipartimento di Fisica ``G. Occhialini'', Universit\`{a} degli Studi di Milano-Bicocca, Piazza della Scienza 3, 20126 Milano, Italy}
\affiliation{INFN, Sezione di Milano-Bicocca, Piazza della Scienza 3, 20126 Milano, Italy}

\author[0000-0002-7616-7136]{Michele Moresco}
\affiliation{Dipartimento di Fisica e Astronomia ``Augusto Righi''--Universit\`{a} di Bologna, via Piero Gobetti 93/2, I-40129 Bologna, Italy}
\affiliation{INAF - Osservatorio di Astrofisica e Scienza dello Spazio di Bologna, via Piero Gobetti 93/3, I-40129 Bologna, Italy}

\author[0009-0003-8886-3184]{Matteo Tagliazucchi}
\affiliation{Dipartimento di Fisica e Astronomia ``Augusto Righi''--Universit\`{a} di Bologna, via Piero Gobetti 93/2, I-40129 Bologna, Italy}
\affiliation{INAF - Osservatorio di Astrofisica e Scienza dello Spazio di Bologna, via Piero Gobetti 93/3, I-40129 Bologna, Italy}

\author[0000-0002-4875-5862]{Francesco Iacovelli}
\affiliation{D\'{e}partement de Physique Th\'{e}orique, Universit\'{e} de Genève, 24 quai Ernest Ansermet, 1211 Genève 4, Switzerland}
\affiliation{Gravitational Wave Science Center (GWSC), Universit\'{e} de Genève, CH-1211 Genève, Switzerland}

\author[0000-0002-4409-5633]{Andrea Cimatti}
\affiliation{Dipartimento di Fisica e Astronomia ``Augusto Righi''--Universit\`{a} di Bologna, via Piero Gobetti 93/2, I-40129 Bologna, Italy}
\affiliation{INAF - Osservatorio Astrofisico di Arcetri, Largo E. Fermi 5, I-50125, Firenze, Italy}

\author[0000-0001-7348-047X]{Michele Maggiore}
\affiliation{D\'{e}partement de Physique Th\'{e}orique, Universit\'{e} de Genève, 24 quai Ernest Ansermet, 1211 Genève 4, Switzerland}
\affiliation{Gravitational Wave Science Center (GWSC), Universit\'{e} de Genève, CH-1211 Genève, Switzerland}

\correspondingauthor{Nicola Borghi}
\email{nicola.borghi6@unibo.it}

\begin{abstract}
\noindent
    With the growing number of gravitational wave (GW) detections and the advent of large galaxy redshift surveys, a new era in cosmology is unfolding. This study explores the synergies between GWs and galaxy surveys to jointly constrain cosmological and astrophysical population parameters. 
    We introduce \href{https://github.com/CosmoStatGW/CHIMERA}{\CHIMERA{} \faGithub}, a novel code for GW cosmology combining information from the population properties of compact binary mergers and galaxy catalogs.
    We study constraints for scenarios representative of the LIGO-Virgo-KAGRA O4 and O5 observing runs, assuming to have a complete catalog of potential host galaxies with either spectroscopic or photometric redshift measurements. 
    We find that a percent-level measurement of $H_0$ could be achieved with the best 100 binary black holes (BBHs) in O5 using a spectroscopic galaxy catalog.
    In this case, the intrinsic correlation that exists between $H_0$ and the BBH population mass scales is broken.
    Instead, by using a photometric catalog the accuracy is degraded up to a factor of $\sim\! 9$, leaving a significant correlation between $H_0$ and the mass scales that must be carefully modeled to avoid bias. 
    Interestingly, we find that using spectroscopic redshift measurements in the O4 configuration yields a better constraint on $H_0$ compared to the O5 configuration with photometric measurements.
    In view of the wealth of GW data that will be available in the future, we argue the importance of obtaining spectroscopic galaxy catalogs to maximize the scientific return of GW cosmology.
\end{abstract}


\section{Introduction} \label{sec:intro}
    Gravitational waves (GWs) emitted by merging compact binaries can serve as \textit{standard sirens} \citep{Schutz1986} as their signal provides a direct measurement of the luminosity distance. This makes GWs a new and powerful cosmological probe for tracing the expansion history of the Universe and testing the validity of general relativity on cosmological scales. First of all, the measurement of the expansion history is a longstanding challenge in the field. 
    The discovery of a discrepancy between the value of the Hubble constant $H_0$ measured from the local distance ladder \citep[e.g.][]{Riess2022} and the one inferred from the cosmic microwave background assuming a $\Lambda$CDM cosmology \citep{PlanckCollaboration2020} motivated a rising interest within the community in developing alternative and emerging cosmological probes \citep[see][for a recent review]{Moresco2022}. 
    Besides being self-calibrated distance tracers, GWs are a particularly promising candidate due to the expected rapid increase in the rate and redshift range of the available catalogs with current and next-generation detectors \citep{Baibhav2019,Iacovelli2022,Branchesi2023}.
    While the measurement of $H_0$ is driven by low-redshift events, those at redshifts $\mathcal{O}(1)$ and beyond open the possibility of testing the distance--redshift relation in full generality. This allows us to constrain the dark energy sector, notably, the phenomenon of \textit{modified GW propagation} \citep[][]{Belgacem2018,Belgacem2018_MG,Amendola2018,Lagos2019}, which is a general prediction of models modifying general relativity at cosmological scales \citep{Belgacem2019}. 

    The peculiarity of GW cosmology is that determining the redshift with GW data alone is not possible because of its inherent degeneracy with binary masses. External information is required to provide cosmological constraints.\footnote{An exception is provided by BNS mergers, which carry a direct redshift signature through the effect of tidal couplings in the inspiral phase \citep{Messenger2012,Messenger2014}. This will be measurable with next-generation ground-based detectors \citep{Dhani2022,Ghosh2022}.}

    A direct measurement of the redshift is possible in the case of a \textit{bright siren}, i.e. an event with the coincident detection of an electromagnetic (EM) counterpart and the consequent identification of the host galaxy which allows the redshift to be directly measured from spectroscopy \citep{Schutz1986,Holz2005,Nissanke2010,Chen2018}.
    Such events are rare, as they typically require mergers involving at least one neutron star and a coincident detection of the EM emission. So far, only the Binary Neutron Star (BNS) event GW170817 has an associated counterpart, the gamma-ray burst GRB 170817A \citep{Abbott2017_GW170817, Abbott2017_MultiM}, which led to the identification of the host galaxy and the first standard siren measurement of $H_0$ \citep{Abbott2017H0}, while being at too low redshift to provide a stringent constraint on modified GW propagation \citep{Belgacem2018_MG}. Detection rates for the ongoing and upcoming observing runs of the LIGO-Virgo-Kagra (LVK) Collaboration are quite uncertain, but in general not optimistic \citep{Colombo2022,Colombo2023}.

    The largest fraction of events is instead made of Binary Black Holes (BBH) without EM counterparts, known as \textit{dark sirens}. Using these sources for cosmology has recently become a concrete option with the latest LVK data \citep{Abbott2021_GWTC2,Abbott2023Catalog}. Two techniques have been primarily employed.
    The first consists of statistically inferring the redshift from potential host galaxies within the GW localization volume \citep{Schutz1986,DelPozzo2012, Chen2018,Gray2020_gwcosmo,Gair2023}.\footnote{A more refined approach involves matching the spatial clustering of GWs and galaxies \citep{Oguri2016,Nair2018,Bera2020,Mukherjee2020,Yu2020,Libanore2021,Mukherjee2021,CigarranDiaz2022,Vijaykumar2023,Bosi2023,Scelfo2023}.}
    This led to measurements of the Hubble constant from LVK data in conjunction with the GLADE~\citep{Dalya2018} and GLADE+~\citep{Dalya2022} galaxy catalogs~\citep{Fishbach2019-DarkGW170817,Abbott2021_O2_H0,Finke2021_DarkSirens,Gray2022,Abbott2023}, as well as the Dark Energy Survey~\citep[DES;][]{SoaresSantos2019,Palmese2020}, the DESI Legacy Survey~\citep{Palmese2023}, the  DECam Local Volume Exploration Survey~\citep{Alfradique2024} and DESI~\citep{Ballard2023}.
    This method also led to the first dark siren bound on modified GW propagation~\citep{Finke2021_DarkSirens}.
    Forecasts indicate that this technique can lead to percent-level measurements with future GW experiments such as LISA~\citep{Laghi2021,Liu2023}, the Einstein Telescope, and Cosmic Explorer~\citep{Muttoni2022,Muttoni2023}.

    Alternatively, the degeneracy between mass and redshift can be broken by modeling intrinsic astrophysical properties as the source-frame mass distribution. In particular, this is possible with \textit{spectral sirens} - sources whose source-frame mass distribution contains features such as breaks, peaks, or changes in slope \citep{Chernoff1993,Taylor2012,Farr2019,Mastrogiovanni2021,Ezquiaga2022}.
    This method allowed in particular to obtain constraints on $H_0$ \citep{Abbott2023} and on modified GW propagation~\citep{Ezquiaga2021_GWTC2,Leyde2022,Mancarella2022,MaganaHernandez2023} from the presence of a mass scale in the BBH distribution \citep{Abbott2021Population,Abbott2023Population}.

    Multiple pipelines have been publicly released, both for the correlation with galaxy catalogs (\texttt{gwcosmo}, \citealt{Gray2020_gwcosmo}; \texttt{DarkSirensStat}, \citealt{Finke2021_DarkSirens}; \texttt{cosmolisa}, \citealt{DelPozzoLaghi2020_cosmolisa}) and for spectral sirens (\texttt{icarogw}, \citealt{Mastrogiovanni2021_icarogw}; \texttt{MGCosmoPop}, \citealt{Mancarella2022}). 
    However, it is known that these two techniques are, in fact, special cases of a unique concept: incorporating prior knowledge about the population in a comprehensive analysis of the ensemble of GW sources \citep{Mastrogiovanni2021,Finke2021_DarkSirens,Moresco2022,Abbott2023}. 
    Since population properties are known in the source frame (mass and redshift), while GW signals are only sensitive to detector-frame quantities (detector-frame mass and luminosity distance), the cosmology-dependent mapping between the two allows breaking the mass-redshift degeneracy statistically.
    In the two aforementioned cases, the population prior used has been either a redshift distribution constructed from a catalog of galaxies, or a model of the source-frame mass distribution. It is natural to seek a unified approach. 
    Moreover, it has been largely recognized that the correlation with a galaxy catalog actually relies on assumptions about the population model for GW sources, and in particular, of a source-frame mass, redshift, and spin distribution \citep{Mastrogiovanni2021,Finke2021_DarkSirens,Abbott2023}. These are needed, in particular, to consistently account for selection effects, i.e. the fact that not all GW sources are equally likely to be observed.
    Moreover, an implicit assumption on the mass, redshift, and spin distributions is inherently present in the form of Bayesian priors adopted in the analysis of individual events, and some assumption has also to be used when accounting for the (in-)completeness of the galaxy catalog. 
    These two facts actually make mandatory a joint analysis of the population properties of compact objects and of the cosmological parameters. Doing so while also incorporating the galaxy catalog information is, however, particularly costly, and until very recently, the population properties of GW sources (and in particular their source-frame mass distribution) have been kept fixed when including the galaxy catalog data.
    
    While this paper was in preparation, two pipelines implementing the joint population and cosmological inference including a galaxy catalog were presented in \cite{Mastrogiovanni2023} and \cite{Gray2023}, namely, \texttt{icarogw2.0} and \texttt{gwcosmo2.0}, building on the homonyms aforementioned codes.
    Their application to the GWTC-3 catalog \citep{Abbott2023Catalog} with the GLADE+ galaxy catalog \citep{Dalya2022} led to updated and more robust measurements of the Hubble constant, as well as of the parameters describing modified GW propagation (the latter being further constrained in \citealt{Chen2023}), and of alternative cosmological models~\citep{Raffai2024}.
    A machine learning-based approach has also been proposed by \cite{Stachurski2023}.
    
    In this paper, we present and release \CHIMERA{} (Combined Hierarchical Inference Model for Electromagnetic and gRavitational wave Analysis), a novel, independent Python code for the joint analysis of GW transient catalogs and galaxy catalogs.\footnote{Available at \href{https://github.com/CosmoStatGW/CHIMERA}{https://github.com/CosmoStatGW/CHIMERA}} \CHIMERA{} is built starting from \texttt{DarkSirensStat}~\citep{Finke2021_DarkSirens} and \texttt{MGCosmoPop}~\citep{Mancarella2022} and allows joint inference of cosmology and population properties of GW sources. This paper is associated with version v1.0.0, which is archived on Zenodo \citep{Borghi2024_CHIMERA}.

    Currently, the information that can be extracted from data with the combined method just described is limited by the lack of a sufficient number of GW events with small enough localization regions and covered by a complete galaxy catalog. In order to fully appreciate the gain achievable from this method, it is therefore of interest to study its potential with upcoming GW data combined with a complete galaxy catalog. This will become a concrete possibility in the coming years, thanks to galaxy surveys already in the data release stage such as DESI~\citep{desi}, and in the data acquisition period such as that of \textit{Euclid}~\citep{Laureijs2011}. In this context, it is also important to understand how large the impact of the redshift measurement uncertainties can be and to assess the relative importance of the information coming from the galaxy survey and from the population model, respectively.
    This is the second goal of the paper, where we explore for the first time the application of this technique in the context of future GW observations and study in detail the impact of different galaxy catalog properties. Here, we focus in particular on the measurement of the Hubble constant.
    We simulate GW detections with detector networks and sensitivities corresponding to the ongoing ``O4'' run of the LVK collaboration, as well as of the future ``O5'' observing run. We study the constraining power of the best 100 events observed in 1~yr, considering a complete galaxy catalog with both spectroscopic and photometric uncertainties on the galaxies' redshifts (for a study of the limited impact of different assumptions on the shape of galaxy redshift uncertainties with current data, see \citealt{Turski2023}).
    We also compare to the case where no information from the galaxy catalog is present.
    
    The paper is organized as follows. In Section~\ref{sec:framework} we present the statistical framework and its implementation in \texttt{CHIMERA}. In Section~\ref{sec:data} we describe the generation of mock galaxy and GW catalogs. Section~\ref{sec:results} provides results for both the \textit{O4-} and \OFive\ scenarios. Finally, Section~\ref{sec:conclusions} concludes the paper and discusses future applications of this method.

\section{Statistical framework} \label{sec:framework}

    We consider a population of GW sources, individually described by the source-frame parameters $\vtheta$, which globally follow a probability distribution $p_\mathrm{pop}(\vtheta \given \vlambda)$ described by hyperparameters $\vlambda$. We summarize here the hierarchical Bayesian inference formalism to infer $\vlambda$ given a catalog of GW detections and a catalog of their potential hosts with measured redshifts.\footnote{For a comprehensive introduction, see \cite{Gair2023}.}
    We separate the hyperparameters describing the underlying cosmology $\vlambda_\mathrm{c}$ (e.g., $H_0$, $\Omega_{\rm m}$) from the ones describing the astrophysical population of GW sources, as the mass distribution $\vlambda_\mathrm{m}$ (e.g., the mass scale, see Section~\ref{sec:mass_dist}) and redshift distribution $\vlambda_\mathrm{z}$ (e.g., the peak of cosmic star formation, see Section~\ref{sec:rate_evol}), so that $\vlambda=\{\vlambda_\mathrm{c}, \vlambda_\mathrm{m}, \vlambda_\mathrm{z}\}$.\footnote{The formalism can be trivially extended to include the spin distribution, which we do not consider in this work.} In this work, $\vlambda_\mathrm{m}$, $\vlambda_\mathrm{z}$ and $\vlambda_\mathrm{c}$ are considered to be independent of each other.
    Given a set $\vdataGW=\{\vdataGW_i\}$ ($i=1,...,N_{\rm ev}$) of data from independent GW events, the population likelihood is \citep{Mandel2019,Vitale2022},
    \begin{equation}\label{eq:hyperlike}
        p(\vdataGW \given \vlambda) \propto \frac{1}{\xi(\vlambda)^{N_{\rm ev}}}\prod_{i=1}^{N_{\rm ev}} \int p(\vdataGW_i \given \vtheta_i, \vlambda_\mathrm{c})\, p_\mathrm{pop}(\vtheta_i \given \vlambda)\, \mathrm{d}\vtheta_i\,,
    \end{equation}
    where $p(\vdataGW_i \given \vtheta_i, \vlambda_\mathrm{c})$ is the individual source likelihood and
    \begin{equation}\label{eq:selection_effects}
        \xi(\vlambda) \equiv \int  P_{\rm det}(\vtheta, \vlambda_\mathrm{c})\,  p_\mathrm{pop}(\vtheta \given \vlambda)\,\mathrm{d}\vtheta\,,
    \end{equation}
    is the selection function, which corrects for selection effects~\citep{Loredo2004,Mandel2019}.
    When $p_\mathrm{pop}(\vtheta \given \vlambda)$ is correctly normalized to unity, $\xi(\vlambda)$ measures the overall fraction of detectable events given $\vlambda$. $P_{\rm det}(\vtheta, \vlambda_\mathrm{c})$ gives the probability of detecting an event with parameters $\vtheta$.
    We consider the following set of source parameters: $\vtheta=\{z, \hat{\Omega}, m_1, m_2\}$, where $z$ is the redshift of the binary, $\hat{\Omega}$ the sky localization, and $m_1$, $m_2$ are the primary and secondary masses.
    
    Note that, when the cosmological hyperparameters $\vlambda_\mathrm{c}$ are included in the analysis, the single-event GW likelihood written in source-frame variables in Eq.~\ref{eq:hyperlike} includes an explicit dependence on $\vlambda_\mathrm{c}$. This is a consequence of the fact that GW observations do not provide information on source-frame parameters, but on detector-frame quantities $\vtheta^\mathrm{det}=\{d_L, \hat{\Omega}, m_1^\mathrm{det}, m_2^\mathrm{det}\}$. These are then mapped to the source frame in a way that depends on the underlying cosmology. 
    Explicitly, the source-to-detector frame conversion requires inverting the relations 
    \begin{align}
        d_L &= d_L(z; \vlambda_{\rm c})\,, \label{eq:dLz}\\
        m_{1,2}^\mathrm{det} &= m_{1,2} \, (1+  z) \label{eq:msourcedet}\, , 
    \end{align}
    where $d_L$ is the luminosity distance, which, in a flat $\Lambda$CDM cosmology, is given by
    \begin{equation}\label{eq:dL_of_z}
        d_L(z; \vlambda_{\rm c}) = c\,(1+z)\int_0^z \frac{\mathrm{d}z'}{H(z'; \vlambda_{\rm c})}\,,
    \end{equation}
    with $\vlambda_{\rm c} \equiv \{H_0,\Omega_{\rm m,0} \}$ and the Hubble parameter being
    \begin{equation}
        H(z; \vlambda_{\rm c}) = H_0 \sqrt{\Omega_{\rm m,0} (1+z)^3+(1-\Omega_{\rm m,0})}\, ,
    \end{equation}
    where the radiation is not considered as its contribution is negligible in the late Universe.
    Alternatively, Eq.~\ref{eq:hyperlike} could be written in the detector frame to remove the dependence on $\vlambda_\mathrm{c}$ (see~\cite{Finke2021_DarkSirens} for a detailed discussion). In this work, for the actual implementation, we find it more convenient to work in source frame, as will be clear below - essentially, this is due to the fact that the galaxy catalog is naturally given in redshift space.
    Note that the single-event GW likelihood is a probability distribution normalized on the data, not on the parameters, so in the transformation from $\vtheta$ to $\vtheta^\mathrm{det}$ it remains unchanged.
    However, in practical applications of hierarchical inference, we are provided with a set of samples drawn from the posterior distribution $p(\vtheta^\mathrm{det}_i\given \vdataGW_i)$ for each event in the GW catalog, obtained with priors $\pi(\vtheta^\mathrm{det}_i)$. 
    It is thus convenient to reexpress the single-event GW likelihood via Bayes' theorem as 
    $p(\vdataGW_i \given \vtheta_i, \vlambda_\mathrm{c}) \propto p(\vtheta_i \given \vdataGW_i, \vlambda_\mathrm{c})/ \pi(\vtheta_i) $. The posterior $p(\vtheta_i \given \vdataGW_i, \vlambda_\mathrm{c})$ can then be obtained from the (cosmology-dependent) conversion of the posterior samples of $p(\vtheta^\mathrm{det}_i\given \vdataGW_i)$ to the source frame, while the prior $\pi(\vtheta_i)$ is related to the one in the detector frame by $\pi(\vtheta_i) = \pi(\vtheta^\mathrm{det}_i) \times | \mathrm{d}\vtheta^\mathrm{det}_i / \mathrm{d}\vtheta_i|$.
    Explicitly, for the variables of interest here, the Jacobian factors are given by
    \begin{equation}\label{eq:jacobians}
        \frac{\mathrm{d} m^\mathrm{det}_{1,2}}{\mathrm{d} m_{1,2}} = 1+z \, , \quad \frac{\mathrm{d}d_L}{\mathrm{d}z}(z;\vlambda_{\rm c}) = \frac{d_L}{1+z} + \frac{c\,(1+z)}{H(z;\vlambda_{\rm c})}\,.
    \end{equation}

    Finally, under the assumption that the mass function does not evolve with cosmic time,\footnote{In this paper, we adopt a time-independent mass distribution. More general models should be envisaged in the future in light of recent results that support evidence of evolution \citep{Rinaldi2023}.} the population function can be split as follows:
    \begin{equation}\label{eq:pop_fcn}
        p_\mathrm{pop}(\vtheta \given \vlambda) = p(m_1, m_2 \given \vlambda_{\rm m})\, p(z, \hat{\Omega} \given \vlambda_{\rm c}, \vlambda_{\rm z})\, ,
    \end{equation}
    where $\hat{\Omega}$ denotes the direction in the sky. In the subsequent sections, we will describe in detail the construction of the population prior taking into account a galaxy catalog and the source-frame mass distribution of the population of compact binaries.

    \subsection{Redshift prior}
    
        The core of the method is the construction of a population prior on the GW source redshift from a galaxy catalog.
        This is given by the second term of Eq.~\ref{eq:pop_fcn} that we factorize as follows:
        \begin{equation}\label{eq:pop_fcn_z_split}
            p(z, \hat{\Omega} \given \vlambda_{\rm c}, \vlambda_{\rm z}) = \frac{p_{\rm gal}(z, \hat{\Omega} \given \vlambda_{\rm c})\, p_{\rm rate}(z \given \vlambda_{\rm z})}{\int p_{\rm gal}(z, \hat{\Omega} \given \vlambda_{\rm c})\, p_{\rm rate}(z \given \vlambda_{\rm z})\,\mathrm{d}z\,\mathrm{d}\hat{\Omega}}
        \end{equation}
        where $p_{\rm gal}$ is the probability that there is a galaxy at ($z, \hat{\Omega}$) and $p_{\rm rate}$ the probability of a galaxy at redshift $z$ to host a GW event. This takes into account that the probability for a galaxy to host a merger can have a nontrivial redshift dependence. We assume that this can be parameterized as 
        \begin{equation}\label{eq:prate}
            p_{\rm rate}(z \given \vlambda_{\rm z}) \propto \frac{\psi(z; \vlambda_{\rm z} )}{(1+z)}
        \end{equation}
        where $\psi(z; \vlambda_{\rm z} )$ is the merger rate evolution of compact objects with redshift and the term $(1+z)^{-1}$ takes into account the conversion between source and detector time. Note that the normalization integral in Eq.~\ref{eq:pop_fcn_z_split}, as well as any rate normalization factor in Eq.~\ref{eq:prate}, simplifies between numerator and denominator in Eq.~\ref{eq:hyperlike} and does not need to be computed explicitly.
        
    \subsection{Galaxy catalog}
    
        As a first approximation, we assume to have a \textit{complete} galaxy catalog, i.e. it contains all the potential host galaxies. In this case, we denote as $p_{\rm cat}(z, \hat{\Omega} \given \vlambda_{\rm c} )$ the probability distribution built from galaxies in the catalog. We can write $p_{\rm gal}(z, \hat{\Omega} \given \vlambda_{\rm c} ) = p_{\rm cat}(z, \hat{\Omega} \given \vlambda_{\rm c} )$
        and compute the latter probability as a sum over the contribution of each galaxy. Given a set of $\vdataEM=\{\vdataEM_g\}$ ($g=1,...,N_{\rm gal}$) EM observations of galaxies we have
        \begin{equation}\label{eq:pcat_complete}
            p_{\rm cat}(z, \hat{\Omega} \given \vlambda_{\rm c} ) =   \frac{\sum_g w_g \, p(z \given \vdataEM_g, \vlambda_{\rm c})\, \delta( \hat{\Omega}-\hat{\Omega}_g) }{\sum_g w_g} 
        \end{equation}
        where $w_g$ weights the probability of each galaxy to host a GW event (e.g., by the galaxy luminosity), $p(z\given \vdataEM_g, \vlambda_{\rm c})$ is the galaxy's redshift distribution that we want to use as a prior, and $\delta( \hat{\Omega}-\hat{\Omega}_g)$ is a Dirac delta distribution of each galaxy's sky localization that can be treated as errorless. 
        
        The galaxy catalog contains redshift measurements $\tilde{z}_g$ and associated uncertainties $\tilde{\sigma}_{z,\,g}$ for each observed galaxy, i.e. $\vdataEM_g=\{\tilde{z}_g,\tilde{\sigma}_{z,\,g}\}$ . From these quantities, we construct the likelihoods, which we assume to be Gaussian, $p(\tilde{z}_g \given z)=\mathcal{N}(z;\tilde{z}_g, \tilde{\sigma}_{z,\,g}^2)$. This is a probability distribution over the observed values $\tilde{z}_g$. To get $p(z \given \vdataEM_g, \vlambda_{\rm c})$ we need to multiply it by a prior on the redshift distribution, which in the absence of other information is naturally chosen as uniform in comoving volume \citep{Gair2023}. Using Bayes' theorem, we get
        \begin{equation}
            p(z \given \vdataEM_g, \vlambda_{\rm c}) = 
            \frac{\mathcal{N}(z; \tilde{z}_g, \tilde{\sigma}_{z,\,g}^2)\, \frac{\mathrm{d}V_c}{\mathrm{d}z}(z;\vlambda_{\rm c}) }{ \int \mathcal{N}(z; \tilde{z}_g, \tilde{\sigma}_{z,\,g}^2)\, \frac{\mathrm{d}V_c}{\mathrm{d}z}(z;\vlambda_{\rm c})\, \mathrm{d}z }\, ,\label{eq:pcat_zprior_onegal}
        \end{equation} 
        where $\mathrm{d}{V_c}/\mathrm{d}{z}$ is the differential comoving volume element in a flat universe. With this definition, Eq.~\ref{eq:pcat_complete} is normalized so that if $p(\tilde{z}_g \given z)=\mathcal{\delta}(z-\tilde{z}_g)$ and in the case of uniform weights, we get $n_{\rm cat}(z) = (1/V_c) \sum_g \delta( z-\tilde{z}_g)$ which is consistent with eq.~3.39 of \cite{Finke2021_DarkSirens}, i.e. the comoving density of galaxies $n_{\rm cat}(z)$ is estimated by counting the objects in the catalog and dividing by the total volume. If, instead, the likelihood is completely uninformative, $p(\tilde{z}_g \given z)=1$, we correctly get a comoving density constant in redshift.
        Note that the assumption of a uniform-in-comoving-volume prior introduces a dependence on the cosmology in the redshift prior, since $\mathrm{d}{V_c}/\mathrm{d}{z}$ is a function of $\vlambda_{\rm c}$. Once normalizing correctly $p(z \given \vdataEM_g, \vlambda_{\rm c})$, however, the dependence on $H_0$ cancels, as this is an overall multiplicative factor in $\mathrm{d}{V_c}/\mathrm{d}{z}$, but a dependence on $\Omega_{\rm m,0}$ remains and should be accounted for. The same would be true if one wished to include more general cosmologies, e.g. a Dark Energy equation of state $w_{\rm DE} \neq -1$. Alternatively, one may adopt a flat prior, which eliminates the extra dependence on the cosmology, as done, e.g. in~\cite{Gray2023}.
            
        In the real world, observed galaxy catalogs are subject to selection effects that result in having an incomplete set of potential hosts. To address this issue, the $p_{\rm gal}$ term in Eq.~\ref{eq:pop_fcn_z_split} has to be modified including the probability of missing galaxies, $p_{\rm miss}(z, \hat{\Omega} \given \vlambda_{\rm c})$, as follows~\citep{Chen2018,Finke2021_DarkSirens}:
        \begin{equation}\label{eq:p_gal}
            p_{\rm gal}(z, \hat{\Omega} \given \vlambda_{\rm c}) = f_\mathcal{R}\, p_{\rm cat}(z, \hat{\Omega} \given \vlambda_{\rm c})+(1-f_\mathcal{R})\, p_{\rm miss}(z, \hat{\Omega} \given \vlambda_{\rm c})\,.
        \end{equation}
        The probability $p_{\rm miss}(z, \hat{\Omega} \given \vlambda_{\rm c})$ is constructed with two pieces of information.
        The first is the number of missing galaxies as a function of redshift and sky position, encoded in the completeness fraction of the catalog $P_{\rm comp}(z, \hat{\Omega})$. This can be estimated with some assumption on the total comoving density of galaxies or on the luminosity distribution of a complete sample that is assumed to follow a Schechter function. We refer to~\cite{Finke2021_DarkSirens} for a rigorous definition of this quantity and for the discussion of different ways to compute it. In \texttt{CHIMERA}, we will use in particular the so-called \textit{mask} completeness of~\cite{Finke2021_DarkSirens}, which we will describe better in Section~\ref{sec:implementation}.
        From this, one can compute the weight $f_\mathcal{R}$ in Eq.~\ref{eq:p_gal} as 
        \begin{equation}\label{eq:fR}
        f_\mathcal{R} \equiv \frac{1}{V_c (\vlambda_{\rm c})}\int P_{\rm comp}(z, \hat{\Omega}) ~\mathrm{d}V_c\,,
        \end{equation}
        where the integral extends to a region sufficiently large to encompass all the GW events under consideration and $V_c (\vlambda_{\rm c})$ denotes the corresponding comoving volume.
        
        The second bit of information amounts to specifying how the missing galaxies are distributed. Two extreme possibilities are a \textit{homogeneous completion}, where the distribution is assumed to be uniform in comoving volume and in sky position, and a \textit{multiplicative completion}, where missing galaxies are assumed to trace the distribution of those present in the catalog~\citep{Finke2021_DarkSirens}. 
        The two correspond to
        \begin{align}
            p_{\rm miss}^{\rm HOM}(z, \hat{\Omega} \given\vlambda_{\rm c}) &= \frac{ 1-P_{\rm comp}(z, \hat{\Omega}) }{ (1-f_\mathcal{R})\,V_c(\vlambda_{\rm c}) } \, \frac{\mathrm{d}V_c}{\mathrm{d}z}(z;\vlambda_{\rm c})\, , \label{eq:pmissHOM} \\
            p_{\rm miss}^{\rm MULT}(z, \hat{\Omega} \given\vlambda_{\rm c}) &= \frac{f_\mathcal{R}}{1-f_\mathcal{R}}\, \frac{1-P_{\rm comp}(z, \hat{\Omega})}{P_{\rm comp}(z, \hat{\Omega})}  \, p_{\rm cat}(z, \hat{\Omega} \given \vlambda_{\rm c})\, .\label{eq:pmissMULT}
        \end{align}
        In practice, as suggested in~\cite{Finke2021_DarkSirens} and done in the code \texttt{DarkSirensStat}, it is convenient to interpolate between a multiplicative completion where the catalog is fairly complete, and a homogeneous completion where the catalog is largely incomplete.
        Finally, an even better option, suggested in~\cite{Finke2021_DarkSirens} and recently studied in detail in~\cite{Dalang2023}, would be to complete the catalog using the typical correlation length of galaxies.

    \subsection{Full Form of the likelihood}\label{sec:fulllike}
    
        By putting together Eqs.~(\ref{eq:hyperlike})--(\ref{eq:fR}) we can write the likelihood in its final form. In the following, we explicitly restrict to the subset of source parameters $\bar{\vtheta}=\{z, \hat{\Omega}, m_1, m_2\}$ and $\bar{\vtheta}^\mathrm{det}=\{d_L, \hat{\Omega}, m_1^\mathrm{det}, m_2^\mathrm{det}\}$. This relies on the assumption that the remaining parameters of the GW waveform (see Eq.~\ref{eq:theta_det} below) have a population distribution that coincides with the prior used in the analysis of individual events. We have
        \begin{widetext}
        \begin{align}
            p(\vdataGW \given \vlambda) &\propto \frac{1}{\xi(\vlambda)^{N_{\rm ev}}} \prod_{i=1}^{N_{\rm ev}} \int \mathrm{d}z\, \mathrm{d}\hat{\Omega} \, \mathcal{K}_{\mathrm{gw},i}(z, \hat{\Omega} \given \vlambda_\mathrm{c}, \vlambda_\mathrm{m}) \,
            p_{\rm gal}(z, \hat{\Omega} \given \vlambda_\mathrm{c})\, \frac{\psi(z ; \vlambda_\mathrm{z})}{1+z}\, , \label{eq:like_full} \\ 
            \mathcal{K}_{\mathrm{gw},i} (z, \hat{\Omega} \given \vlambda_\mathrm{c}, \vlambda_\mathrm{m}) &\equiv 
            \int \mathrm{d}m_1 \mathrm{d}m_2 \, \frac{ p(z, m_1, m_2, \hat{\Omega} \given \vdataGW_i, \vlambda_\mathrm{c})}{ \pi( d_L ) \pi( m_1^\mathrm{det} ) \pi( m_2^\mathrm{det} ) } \, \frac{1}{\frac{\mathrm{d} d_L}{\mathrm{d} z}(z; \vlambda_\mathrm{c})\, (1+z)^2}\,  p(m_1, m_2 \given \vlambda_\mathrm{m})\,,\label{eq:Kgw}\\
            \xi(\vlambda) & = \int \mathrm{d} \vtheta \, P_{\rm det}(\vtheta, \vlambda_\mathrm{c})\, \,  p(m_1, m_2 \given \vlambda_\mathrm{m}) \, p_{\rm gal}(z, \hat{\Omega} \given \vlambda_\mathrm{c})\, \frac{\psi(z ; \vlambda_\mathrm{z})}{1+z} \, . \label{eq:selection_effects_xi} 
        \end{align}
        \end{widetext}

        Note that we have omitted the overall normalization of the redshift prior, as this is independent of $z, \hat{\Omega}$ and is present also in the selection function, leading to a simplification between numerator and denominator in Eq.~\ref{eq:like_full}. A different normalization of this quantity just amounts to a different interpretation of the selection function, which does not correspond to the fraction of detectable events, as a consequence of the fact that $p_\mathrm{pop}$ loses its correct normalization to unity.
        We find the form of Eqs.~(\ref{eq:like_full})--(\ref{eq:selection_effects_xi}) useful since it makes clear what quantities actually need to be computed in the numerical evaluation. On the other hand, we note that the normalization of the term $p_{\rm gal}$ has to be computed carefully, in order to correctly balance the contributions of the catalog and missing terms in Eq.~\ref{eq:p_gal}.
        
        \begin{figure}
            \centering
            \includegraphics[width=0.449\textwidth]{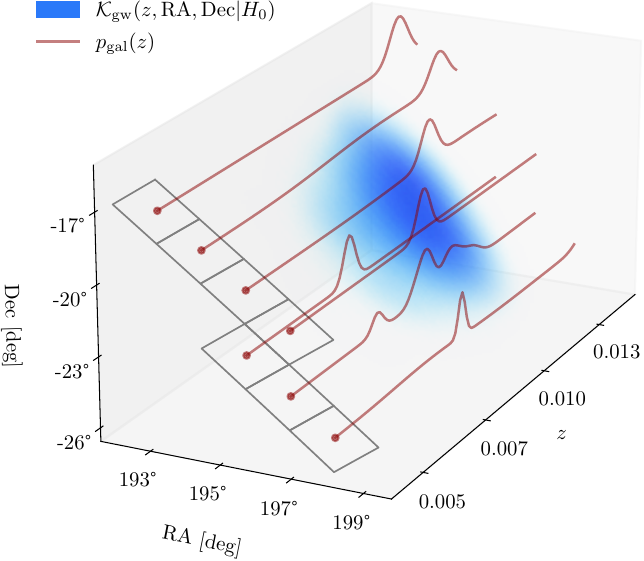}
            \caption{Visual representation of the underlying workflow of \CHIMERA{}. The GW probability is approximated by a three-dimensional KDE $\mathcal{K}_\mathrm{gw}$, while the galaxy probability $p_\mathrm{gal}$ is evaluated by summing the individual galaxy contributions within the pixel enclosed in the GW sky localization area.}
            \label{fig:CHIMERA_sketch}
        \end{figure}
        
        \begin{figure*}[ht]
            \centering
            \includegraphics[width=0.99\textwidth]{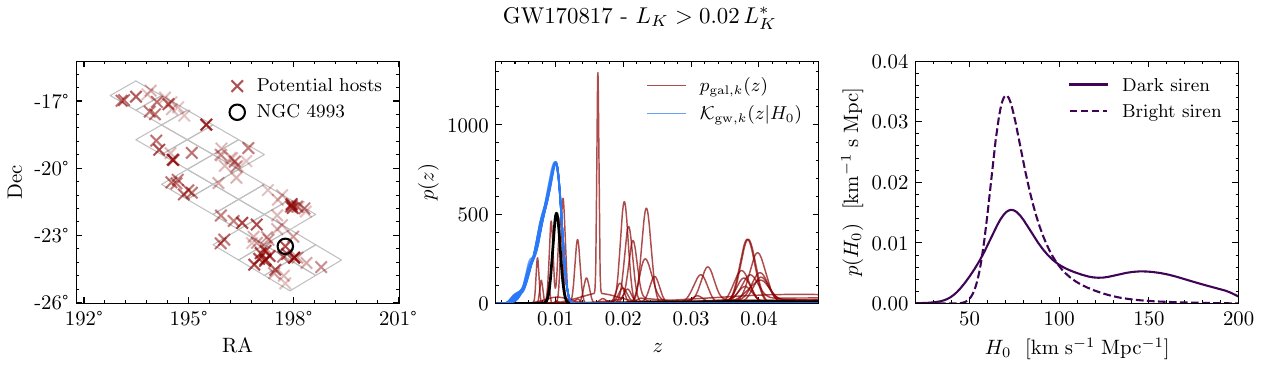}
            \caption{Analysis of GW170817 with \texttt{CHIMERA} using GLADE+ galaxies with a luminosity $L_K>0.02\,L_K^\ast$. \textit{Left}: sky distribution of the potential host galaxies (red crosses) contained in the 90\% credible pixels (gray squares). The true host NGC~4993 is identified with a gold circle. \textit{Center:} redshift distribution of the GW kernel (blue) assuming $H_0=70~\kmsMpc$ and galaxy probability $p_\mathrm{gal}$ (red) in each pixel. The pixel including NGC~4993 is colored in black. \textit{Right}: posterior distributions on $H_0$ assuming that the true host is or is not observed.}
            \label{fig:GW170817}
        \end{figure*}

    \subsection{Implementation in \CHIMERA{}}\label{sec:implementation}
    
        We now discuss the implementation of the method in \CHIMERA{}. The code accurately computes Eq.~\ref{eq:like_full} across different regimes, i.e. when the EM redshift information in highly constraining (bright sirens), more loosely constraining (dark sirens), or not present at all (spectral sirens). We also prioritize computational efficiency to meet the demands of next-generation GW observatories and upcoming galaxy surveys. 
        
        In this section, we summarize the key aspects of the code, while a more detailed description is provided in Appendix~\ref{app:code}. 
        We will use the real event GW170817 as an illustrative example, considering it a dark siren in conjunction with the GLADE+ galaxy catalog. This is useful because GW170817 is the best-localized GW event so far and its host galaxy, NGC~4993, has been uniquely identified~\citep{Abbott2017_GW170817}, providing an ideal benchmark for dark siren techniques. Indeed, this is often used as a test case and allows comparison to other pipelines~\citep{Mastrogiovanni2023,Gray2023}.
        In Appendix~\ref{app:GW170817}, we provide details about the data and cuts used for this example. 
        
        Figure~\ref{fig:CHIMERA_sketch} shows a compact representation of the workflow of \CHIMERA{}, in the case of GW170817. In short, the galaxy catalog term is precomputed as a function of redshift (red lines) on a pixel-by-pixel basis inside the GW localization region (given by the gray squares). The GW kernel in Eq.~\ref{eq:Kgw} is given by a kernel density estimation (KDE) of the GW posterior samples re-weighted for the population prior on source-frame masses (blue area).

        \paragraph{Pixelization and precomputation of the galaxy term} 
        The quantity $p_{\rm gal}(z,\hat{\Omega} \given \vlambda_{\rm c})$ can be cosmology dependent, but only when adopting a cosmology-dependent prior on the galaxies' redshifts, and in any case never on $H_0$, as explained in Section~\ref{sec:framework}. If adopting a flat prior on the galaxies' redshifts, or a narrow prior on the other cosmological parameters (e.g. a Planck prior on the matter density), the quantity $p_{\rm gal}$ can be computed once for all, given a specific galaxy catalog. 
        For each event in the GW catalog, \CHIMERA{} selects galaxies enclosed in the GW localization volume, with a redshift range computed taking into account the whole prior range for $H_0$. Then, the sky area of the GW event is divided into equal-area pixels in the pixelization scheme of \texttt{Healpix} \citep{Gorski2005,Zonca2019}. \CHIMERA{} includes an adaptive pixelization procedure that allows us to set the pixel size for each GW event depending on its localization area (see App.~\ref{app:code} for details).
        This step corresponds to the left panel of Fig.~\ref{fig:GW170817}, where we show the galaxies enclosed in the $90\%$ localization volume of GW170817 as red crosses, overlapping the pixels.
        The quantity $p_{\rm gal}(z,\hat{\Omega} \given \vlambda_{\rm c})$ is then approximated as a redshift-dependent function of the form given in Eqs.~(\ref{eq:pcat_complete})--(\ref{eq:pcat_zprior_onegal}) for each pixel, by summing the contribution of all the galaxies enclosed within it. 
        The weights in Eq.~\ref{eq:pcat_complete} can be chosen to be either uniform or corresponding to the galaxies' luminosity in the catalog.
        The redshift distributions in each pixel are shown in red in the central panel of Fig.~\ref{fig:GW170817} for the example of GW170817.
        This method allows for the compression of the information coming from all galaxies in each pixel in a single function of redshift. In the case of GW170817, we find 88 galaxies in the 90\% localization volume, while using 20 pixels, which corresponds to a factor of $\sim 5$ improvement in the computational cost. This procedure is even more effective in the case of events with thousands of galaxies in the localization region.
        We note that, in general, the GW skymap changes on much larger angular scales than the galaxy distribution. This means that the compression introduced here (which basically amounts to approximating the galaxies' angular positions with the center of each pixel) does not lead to a loss of information as far as the pixel size is chosen appropriately.

        \paragraph{Catalog completeness} The completeness fraction $P_{\rm comp}(z, \hat\Omega)$ is computed following the mask method presented in \cite{Finke2021_DarkSirens}. First, similar pixels are grouped together in masks by applying an agglomerative clustering algorithm. The chosen feature for this purpose is the number of galaxies within each pixel. The number of masks is arbitrarily chosen depending on the survey properties. For each mask, the completeness fraction is computed by comparing the luminosity function at each redshift to a reference Schechter function. We refer to \cite{Finke2021_DarkSirens} for a more detailed description of this procedure. In particular, this allows us to easily cut regions particularly under-covered by the galaxy survey, for example, the region of the sky covered by the  Milky Way. It also makes straightforward the use of galaxy catalogs with partial sky coverage. In the example of GW170817, it is complete.
    
        \paragraph{3D GW kernel} We now turn to the term $\mathcal{K}_\mathrm{gw}$ in Eq.~\ref{eq:Kgw}. For each GW event, we have a set of posterior samples of the detector-frame quantities  $\bar{\vtheta}^\mathrm{det}=\{d_L, \hat{\Omega}, m_1^\mathrm{det}, m_2^\mathrm{det}\}$ approximating the posterior probability $p(m_1^\mathrm{det}, m_2^\mathrm{det}, d_L, \hat{\Omega}\given \vdataGW_i)$. \CHIMERA{} assumes that the prior on detector-frame masses used for the individual event analysis is flat, $\pi( m_{1}^\mathrm{det})=\pi( m_{2}^\mathrm{det})=1$, while the prior on the luminosity distance is $\pi(d_L)\propto d_L^ 2$. Then, for each value of $\vlambda$, the samples are converted to source-frame quantities by inverting Eqs.~(\ref{eq:dLz})--(\ref{eq:dL_of_z}). Each sample, labeled  by $j$, is assigned a weight equal to 
        \begin{equation}
            w_{\mathcal{K},j}(\vlambda) =  \frac{p(m_{1,j}, m_{2,j} \given \vlambda_{\rm m})}{d_L^ 2\,\frac{\mathrm{d} d_L}{\mathrm{d}z}(z_j; \vlambda_\mathrm{c})\, \left(1+z_j\right)^2}   \, ,
        \end{equation}
        and $\mathcal{K}_\mathrm{gw}$ is obtained by a 3D weighted KDE in the space $(z, \mathrm{RA}, \mathrm{Dec})$ with weights $w_{\mathcal{K},j}(\vlambda)$. Using the re-weighted samples in this subspace corresponds to performing the integral in Eq.~\ref{eq:Kgw} with the correct population prior $p(m_{1}, m_{2} \given \vlambda_{\rm m})$ on the \emph{source-frame} masses. This will encode a dependence on the cosmological parameters, which enters in the conversion of the posterior samples from detector frame to source frame, resulting in a reshaping of the kernel $\mathcal{K}_\mathrm{gw}$. A non-flat source-frame mass spectrum will add redshift information at the statistical level through this term, even in absence of a galaxy catalog, in which case we recover the spectral siren case.
        For GW170817, the GW kernel is plotted in blue in Fig.~\ref{fig:GW170817}~(central panel), assuming a value of $H_0=70~\kmsMpc$ and a flat mass distribution between $1$ and $3~\msun$.
    
        \paragraph{Integration} The double integral in Eq.~\ref{eq:like_full} is performed by integrating numerically $\mathrm{d}z$ in each pixel first. Then, the integral in $\mathrm{d}\hat{\Omega}$ can be approximated as a discrete sum of the results in each pixel, multiplied by the pixel area. The latter, for each given GW event, is equal for all pixels and inversely proportional to the number of Healpix pixels of the map. So, in the end, the angular integral in Eq.~\ref{eq:like_full} is obtained as an average over pixels.
    
        \paragraph{Selection effects} The selection bias term $\xi(\vlambda)$ from Eq.~\ref{eq:selection_effects} is computed by using the standard re-weighted Monte Carlo (MC) method~\citep{Tiwari2018,Farr2019_Neff}. In this case, it is crucial to use \emph{detector frame} quantities, since the detectability in a GW experiment is a function of those only, i.e. $P_{\rm det} = P_{\rm det}(\vtheta^{\rm det})$. This allows us to compute this function only once. 
        The integral in Eq.~\ref{eq:selection_effects_xi} can then be computed with the change of variables $\vtheta \mapsto \vtheta^{\rm det}(\vtheta, \vlambda_\mathrm{c})$, where the Jacobian factors are given in Eq.~\ref{eq:jacobians}.
        \CHIMERA{} takes as input a set of simulated events, each with detector-frame parameters $\vtheta_i^{\rm det}$, drawn from a reference distribution with probability $p_\mathrm{draw}(\vtheta_i^{\rm det})$. These must have been previously injected in the same detection pipeline used to obtain the GW catalog, storing those that pass the detection threshold. 
        
        Then, the selection function in Eq.~\ref{eq:selection_effects_xi} is computed with MC integration as 
        \begin{equation}\label{eq:bias}
        \begin{split}
            \xi(\vlambda)=&\frac{1}{N_\mathrm{inj}} \sum_{i=1}^{N_\mathrm{det}} \frac{1}{p_\mathrm{draw}(\vtheta_i^{\rm det})}  \frac{1}{\frac{\mathrm{d}d_L}{\mathrm{d}z}(z_i, \vlambda_\mathrm{c}) (1+z_i)^2}  \\
            & \times p(m_{1,i}, m_{2,i} \given \vlambda_\mathrm{m}) \, p_{\rm gal}(z_i, \hat{\Omega}_i \given \vlambda_{\rm c})\, \frac{\psi(z_i ; \vlambda_{\rm z})}{1+z_i}
        \end{split}
        \end{equation}
        where $N_\mathrm{inj}$ is the total number of injections, $N_\mathrm{det}$ the number of detected ones, and the source-frame quantities $z_i$, $m_{1,i}$, and $ m_{2,i}$ are understood to be computed from the (fixed) detector-frame ones via inversion of Eqs.~(\ref{eq:dLz})--(\ref{eq:dL_of_z}), and are thus functions of the cosmological parameters $\vlambda_\mathrm{c}$.
        Finally, when computing the selection bias term, \CHIMERA{} employs a galaxy catalog interpolant $p_{\rm gal}$ constructed on the full sky.

        \paragraph{Accuracy of the MC integrals}  Finally, when computing Eq.~\ref{eq:bias}, we ensure that the effective number of independent draws after re-weighting, $N_\mathrm{eff}$, for a given population is sufficiently high ($N_\mathrm{eff}>5N_\mathrm{det}$, see \citealt{Farr2019_Neff}). This criterion is also applied to the GW kernel weights to ensure numerical stability.  \\
    
    In conclusion, the right panel of Fig.~\ref{fig:GW170817} shows the results for the posterior probability on $H_0$ (solid line). We obtain a value of $H_0=73^{+58}_{-22}~\kmsMpc$ in the dark siren case, in good agreement with \cite{Fishbach2019-DarkGW170817}. As can be seen from Fig.~\ref{fig:GW170817} (central panel), the primary contribution to the posterior arises from galaxies at approximately $z\sim 0.01$. This value combined with the GW measurement of $d_L\sim40$~Mpc, implies $H_0\sim 70~\kmsMpc$. The galaxy groups at $z\sim 0.02~(0.04)$ provide a shallower contribution to $H_0\sim 150~(300)~\kmsMpc$ due to the presence of selection effects that disfavor high values of $H_0$ [$\xi(H_0)\sim H_0^3$]. In this case, the assumption of a population model for GW170817 provides no significant improvements to $H_0$.
    
    For comparison and to test the performance of the code in the bright siren regime, we repeat the analysis assuming the identification of the host galaxy with $z=0.0100\pm0.0005$ \citep{Abbott2017_GW170817}. In this case, we assign a weight $w=1$ to NGC~4993 and $w=0$ to all the other galaxies (see Eq.~\ref{eq:pcat_complete}). The resulting posterior on $H_0$ is shown in Fig.~\ref{fig:GW170817}c (dashed line). In the bright siren case, we obtain a value of $H_0=69^{+15}_{-8}~\kmsMpc$, in very good agreement with \cite{Abbott2023}.

\section{Mock Catalogs}\label{sec:data}

    In this section, we describe the procedure to build realistic mock galaxy and GW catalogs. These will be used to both validate the code and provide updated forecasts for \textit{O4-} and \OFive\ detector configurations.

    \subsection{Parent Galaxy Sample}

        We generate our mock galaxy catalog (hereafter, \textit{parent sample}) from the MICE Grand Challenge light-cone simulation (v2), which populates one octant of the sky (close to 5157 $\mathrm{deg^2}$) designed to mimic a complete DES-like (Dark Energy Survey) survey up to an observed magnitude of $i<24$ at redshift $z<1.4$~\citep{Carretero2015,Crocce2015,Fosalba2015a,Fosalba2015b,Hoffmann2015}. MICE assumes a flat $\Lambda$CDM cosmology with $H_0=70~\kmsMpc$, $\Omega_{m,0}=0.25$, and $\Omega_{\Lambda,0}=0.75$.

        While we ideally require a complete catalog with high number density, as a simplifying assumption in this paper we consider only galaxies with stellar masses $\log M_\star/\msun > 10.5$. This cut is consistent with the idea that the binary merger rate is traced by stellar mass, as also adopted in current standard sirens analysis via absolute magnitude cuts and luminosity weighting \citep{Fishbach2019-DarkGW170817,Finke2021_DarkSirens,Gray2022,Abbott2023,Mastrogiovanni2023,Gray2023}. A similar cut in mass is also considered in the context of simulations for the Einstein Telescope \citep[e.g.,][]{Muttoni2023}. 

        We subsample the MICEv2 catalog to reproduce the density for the cut described above extracting the galaxies to get a uniform in comoving volume distribution. In the end, we obtain a parent sample of about 1.6 million massive galaxies.

        For the redshift uncertainties, we consider two cases. First of all, we explore the possibility of maximizing the galaxy catalog information by having a spectroscopic catalog. This could be done by expanding the currently available catalogs \citep[GLADE+][]{Dalya2022} in the future by exploiting the information provided by the next large spectroscopic surveys. As an example, the ESA mission Euclid \citep{Laureijs2011} will provide an all-sky map of spectroscopic redshift in the range $0.9<z<1.8$, with an accuracy of $\sigma_z/(1+z)\lesssim0.001$, and the Dark Energy Spectroscopic Instrument \citep[DESI][]{desi} is planned to observe $\sim 14,000$ deg\textsuperscript{2} covering the redshift range of $0.4<z<2.1$.
        Second, we study how the information extracted changes when using photometric redshift, assuming an uncertainty $\sigma_z/(1+z)=0.05$. This is easily accessible with current ongoing surveys like DES (e.g., DES has reached $\sigma_z\sim 0.01$, \citealt{DES_photoz}, and this limit can be pushed to $\sigma_z\sim 0.007$ with improved techniques, \citealt{DES_phenoz}) on a smaller area, and in future surveys like Euclid and Rubin Observatory \citep{rubin} are planned to extend it to the entire sky to a depth of Euclid H magnitude $H_E\sim24$ ($H_E\sim26$ in the Deep Survey) and an expected uncertainty $\sigma_z/(1+z)\lesssim 0.05$ \citep{Desprez2020,Schirmer2022}.

        Therefore, in this paper, we consider two regimes of photometric (hereafter, ``phot'') and spectroscopic  (hereafter, ``spec'') redshift uncertainties:
        \begin{equation}\label{eq:zerr:def}
            \sigma_z = \begin{cases}
                  0.001\,(1+z) & (\zs) \\
                  0.05\,(1+z)  & (\zp) 
            \end{cases}
        \end{equation}

        \begin{deluxetable*}{llcc}[t]
        \tablewidth{0.99\textwidth}
        \caption{Summary of the Population Hyperparameters and Priors Adopted. $\mathcal{U(\cdot)}$ denotes a uniform distribution.}\label{tab:parameters}
        \tablehead{\colhead{Parameter} & \colhead{Description} & \colhead{Fiducial Value} & \colhead{Prior}}
        \startdata
            & \textbf{Cosmology (flat $\Lambda$CDM)} \\
            $H_0$ & Hubble constant $[\kmsMpc]$ & 70.0 & $\mathcal{U}({10.0}, {200.0})$ \\
            $\Omega_{\rm m,0}$ & Matter energy density & 0.25 & Fixed \\
            \hline
            & \textbf{Rate evolution (Madau-like)} \\
            $\gamma$ & Slope at $z<z_p$ & 2.7 & $\mathcal{U}({0.0}, {12.0})$ \\
            $\kappa$ & Slope at $z>z_p$ & 3 & $\mathcal{U}({0.0}, {6.0})$ \\
            $z_{\rm p}$ & Peak redshift & 2 & $\mathcal{U}({0.0}, {4.0})$ \\
            \hline
            & \textbf{Mass distribution (PowerLaw+Peak)} \\
            $\alpha$ & (Primary) slope of the power law & 3.4 & $\mathcal{U}({1.5}, {12.0})$ \\
            $\beta$ & (Secondary) slope of the power law & 1.1 & $\mathcal{U}({-4.0}, {12.0})$ \\
            $\delta_m$ & (Primary) smoothing parameter [$\msun$] & 4.8 & $\mathcal{U}({0.01}, {10.0})$ \\
            $m_{\rm low}$ & Lower value $[\msun]$& 5.1 & $\mathcal{U}({2.0}, {50.0})$ \\
            $m_{\rm high}$ & Upper value $[\msun]$& 87.0 & $\mathcal{U}({50.0}, {200.0})$ \\
            $\mu_{\rm g}$ & (Primary): mean of the Gaussian component $[\msun]$ & 34.0 & $\mathcal{U}({2.0}, {50.0})$ \\
            $\sigma_{\rm g}$ & (Primary): standard deviation of the Gaussian component $[\msun]$ & 3.6 & $\mathcal{U}({0.4}, {10.0})$ \\
            $\lambda_{\rm g}$ & (Primary): fraction of the Gaussian component  & 0.039 & $\mathcal{U}({0.01}, {0.99})$
        \enddata
        \end{deluxetable*}

        \begin{figure*}[ht]
            \includegraphics[width=0.976\textwidth]{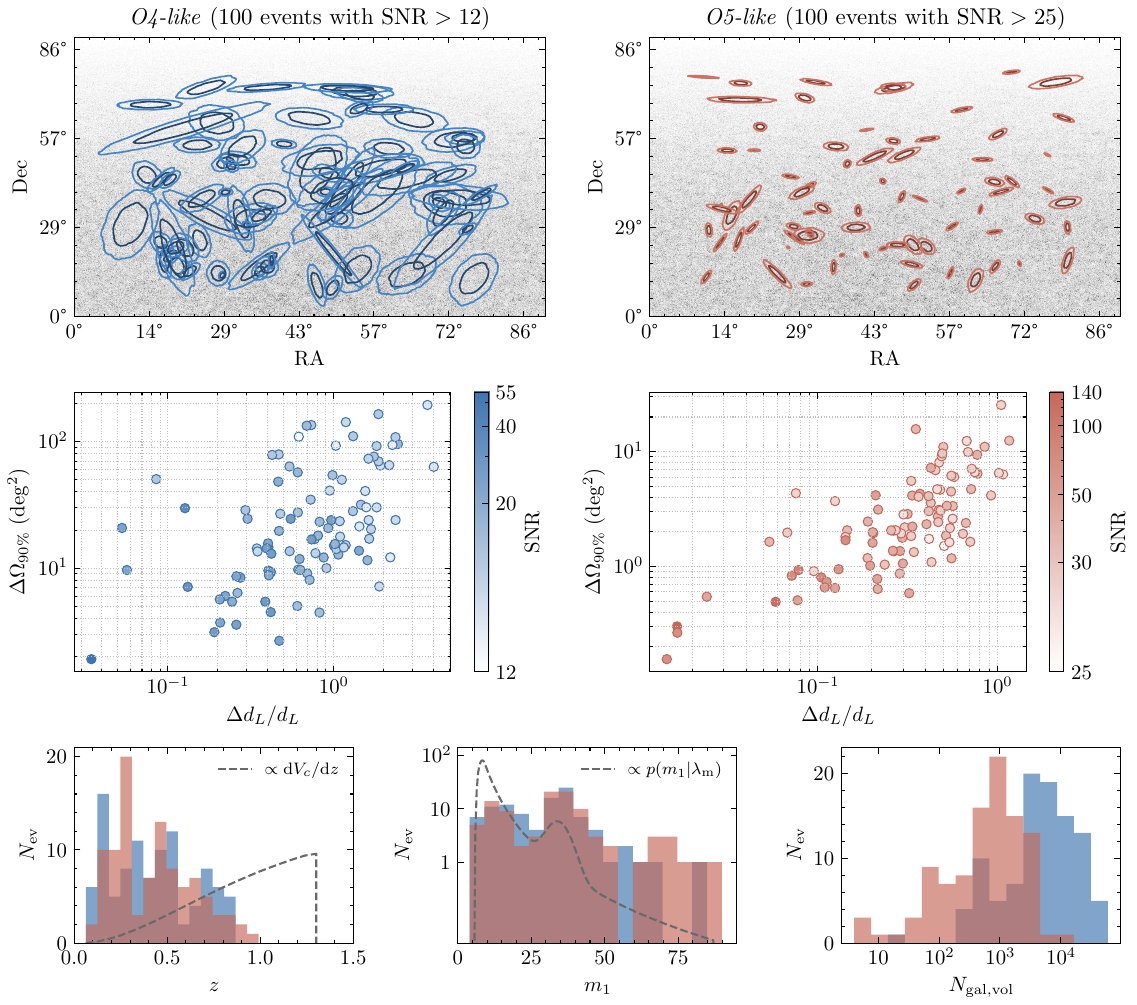}
            \caption{Main properties of the simulated \OFour\ (blue) and \OFive\ (red) GW catalogs. \textit{Upper panels:} GW sky localization areas at $1$ and $2\,\sigma$ overlaid to their potential host galaxies (gray) extracted from MICEv2. \textit{Middle panels:} distribution of the relative uncertainty on the luminosity distance and sky localization area as a function of network S/N. \textit{Lower panels:} distribution of the GW events as a function of  (\textit{left to right}): redshift, primary mass, and number of galaxies contained in the GW localization volume.
            }\label{fig:sampleGW}
        \end{figure*}
    
    \subsection{Sample of GW events}\label{sec:GWsample}
    
        We generate mock GW events from the parent sample by fixing cosmological hyperparameters $\vlambda_{\rm c}$ and astrophysical population hyperparameters $\vlambda_{\rm z}$ and $\vlambda_{\rm m}$. We describe the redshift and mass distributions in turn.
        
        \subsubsection{Rate evolution}\label{sec:rate_evol}
            The source-frame merger rate is parameterized as follows \citep[see][]{Madau2014}:
            \begin{equation}\label{eq:rate_MD}
                \psi(z ; \vlambda_{\rm z}) = \frac{(1+z)^{\gamma}}{1 + \left (  \frac{1+z}{1+z_\mathrm{p}}   \right )^{\gamma + \kappa}},
            \end{equation}
            where $\psi(z)\propto (1+z)^\gamma$ at low $z$, then reaches its peak near $z_\mathrm{p}$, and subsequently declines as $\psi(z)\propto(1+z)^{-\kappa}$. To build the catalog, we use $\gamma=2.7$ consistent with the LVK GWTC-3 results \citep{Abbott2023Population}. The limited detection range of current GW detectors restricts our ability to determine the merger rate at higher redshifts, therefore we adopt $z_\mathrm{p}=2$, $\kappa=3$, consistent with the idea that $\psi(z ; \vlambda_{\rm z})$ follows the galaxy's star formation rate density with parameters from \cite{Madau2014}. The catalog of potential sources is then obtained by sampling the parent sample using a weight proportional to the detector-frame merger rate $\psi(z ; \vlambda_{\rm z})/(1+z)$.

        \subsubsection{Mass distribution}\label{sec:mass_dist}
            For the mass distribution, we adopt the phenomenological ``PowerLaw+Peak'' (PLP) model following the LVK GWTC-3 results~\citep{Abbott2023Population}.
            The mass term of Eq.~\ref{eq:pop_fcn} is factorized as follows: 
            \begin{equation}\label{eq:pop_fcn_mass_split}
                p(m_1, m_2 \given \vlambda_{\rm m}) = p(m_1\given \vlambda_{\rm m})\, p(m_2\given m_1,\vlambda_{\rm m}) \,.
            \end{equation}
            The probability of the primary BH mass is given by
            \begin{equation}
                p(m_1\given \vlambda_{\rm m}) \propto \left[(1 - \lambda_{\rm g})\,\mathcal{P}(m_1) + \lambda_{\rm g}\, \mathcal{G}(m_1)\right]\, \mathcal{S}(m_1) \,,
            \end{equation}
            where $\mathcal{P}(m)\propto m^{-\alpha}$ is a power law truncated in the domain $m\in[m_{\rm low},m_{\rm high}]$, $\mathcal{G}(m)\propto \mathcal{N}(\mu_\mathrm{g}; \sigma_\mathrm{g}^2)$ is a Gaussian component whose contribution relative to $\mathcal{P}$ is regulated by $\lambda_{\rm g}$, and $\mathcal{S}(m)\in[0,1]$ is a smoothing piece-wise function with a tapering parameter $\delta_m$ fully described in appendix B of \cite{Abbott2021Population}. The secondary BH mass is modeled by a power law with an index $\beta$ in the domain $m\in[m_{\rm low},m_1]$.

        \subsubsection{Summary}
            With the above assumptions, the cosmological and astrophysical hyperparameters to be studied are
            \begin{equation}\label{eq:allparams}
            \begin{split}
                    \vlambda_{\rm c} &= \{H_0, \Omega_{\rm m,0}\} \\
                    \vlambda_{\rm z} &= \{\gamma, k, z_{\rm p} \} \\
                    \vlambda_{\rm m} &= \{\alpha, \beta, \delta_m, m_{\rm low}, m_{\rm high}, \mu_{\rm g}, \sigma_{\rm g}, \lambda_{\rm g} \}.
            \end{split}
            \end{equation}
            The fiducial values and the prior ranges chosen for this work are reported in Table~\ref{tab:parameters}. For the cosmological parameters, we will consider the value of the matter density to be fixed to its fiducial value.

    \subsection{GW data generation}
    
        For the simulation of GW detections, we use the simulation pipeline \texttt{GWFAST}\footnote{Available at \url{https://gwfast.readthedocs.io/en/latest/}} \citep{Iacovelli2022,Iacovelli2022-GWFAST}. We assume quasi-circular non-precessing BBH systems. Their waveform is characterized by the detector-frame parameters
        \begin{equation}\label{eq:theta_det}
            \vtheta^\mathrm{det}=\left\{\mathcal{M}_c, \eta, \chi_{1, z}, \chi_{2, z}, d_L, \theta, \phi, \iota, \psi, t_c, \Phi_c\right\}\,,
        \end{equation}
        where $\mathcal{M}_c$ is the detector-frame chirp mass, $\eta$ is the symmetric mass ratio, $\chi_{1/2, z}$ are the adimensional spin parameters along the direction of the orbital angular momentum, $d_L$ is the luminosity distance, $\theta = \pi/2 - \mathrm{Dec}$ and $\phi = \mathrm{RA}$ are the sky position angles, $\iota$ refers to the inclination angle of the binary's orbital angular momentum with respect to the line of sight, $\psi$ is the polarization angle, $t_c$ is the coalescence time, and $\Phi_c$ is the phase at coalescence.

        First of all, we generate a population of GW events following the prescriptions given in Section~\ref{sec:GWsample}, so that each source is characterized by a set of source-frame parameters $\vtheta$. For the parameters in Eq.~\ref{eq:theta_det} that are not explicitly mentioned in Section~\ref{sec:GWsample}, we assume the following distributions: the sky position angles are uniform on the octant covered by the parent sample, the inclination angles have a uniform distribution in $\cos\iota$ in the range $[0,\,\pi]$, the polarization angle and coalescence phase have a uniform distribution in $[0,\,\pi]$ and $[0,\,2\pi]$, respectively, and the time of coalescence, given as Greenwich mean sidereal time and expressed in units of fraction of a day, is uniform in the interval $[0,\,1]$. For the spin parameters instead, we adopt a flat distribution between $[-1,\,1]$ for the components aligned with the orbital angular momentum.
        The source-frame parameters are then converted to detector frame using the fiducial cosmology (see Table~\ref{tab:parameters}).
        
        For each source, we simulate GW emission and compute the network matched-filtered signal-to-noise ratio ($\mathrm{S/N}$) using the \texttt{IMRPhenomHM} \citep{London2018,Kalaghatgi2020} waveform approximant, which includes the contribution of subdominant modes to the signal, that are of great relevance to describe in particular the merger phase of BBH systems. We consider two network configurations.
        The first, denoted as \OFour, is composed of a network of the two LIGO interferometers at Hanford and Livingston, USA, the Virgo interferometer in Cascina, Italy, and the KAGRA interferometer in Japan. For the second configuration, denoted as \OFive\, the network includes the two LIGO, Virgo, and KAGRA instruments, as well as a LIGO detector located in India. We assume sensitivity curves representative of the O4 and O5 runs of the LVK Collaboration, with public sensitivity curves from~\cite{Abbott2016}.\footnote{The amplitude spectral densities can be found at~\url{https://dcc.ligo.org/LIGO-T2000012/public}. For O4, we use \texttt{kagra\_3Mpc} for KAGRA. For O5, we use \texttt{AplusDesign} for the three LIGO detectors, \texttt{avirgo\_O5low\_NEW} for Virgo, and \texttt{kagra\_80Mpc} for KAGRA.} We assume a $100\%$ duty cycle in all cases. 

        Then, we select samples of GW events as follows:
        \begin{itemize}
            \item \OFour: 100 events with a network $\mathrm{S/N}>12$;
            \item \OFive: 100 events with a network $\mathrm{S/N}>25$.
        \end{itemize}

        These cuts are designed to yield the 100 best events for each configuration over approximately 1~yr of observation. We determined these numbers simulating with \texttt{GWFAST} a 1~yr observing run for each of the two scenarios, with a population of BBHs with an overall merger rate calibrated on the latest LVK constraints \citep[see, e.g.][]{Iacovelli2022}. We note that the fact that the simulation is performed on one octant of the sky is irrelevant in our case, as we do not constrain direction-dependent hyperparameters.
        
        We approximate the GW posterior probability with the Fisher Information Matrix (FIM) approximation, where the GW likelihood $p(\vdataGW_i \given \vtheta_i^{\rm det})$ is assumed to be given by a multivariate Gaussian distribution. This approximation is valid for the high S/N events that we are considering in this work (see~\cite{Iacovelli2022} for details). We compute the FIM with \texttt{GWFAST}. For each event, we then draw $5000$ samples from a multivariate Gaussian distribution with covariance equal to the inverse of the FIM, further imposing priors that are chosen to be flat in detector-frame masses with the condition $m_2^{\rm det}<m_1^{\rm det}$, $\propto d_L^2$ for the luminosity distance, and equal to the distributions used to sample the events for the remaining parameters. We then use the samples of $m_1^{\rm det}, m_2^{\rm det}, d_L, \mathrm{RA}, \mathrm{Dec}$ to compute the GW likelihood as described in Section~\ref{sec:implementation}.
        
        The main properties of the galaxy and GW catalogs are summarized in Fig.~\ref{fig:sampleGW}. The top panels show the GW skymaps of the events detected in the two configurations overlaid to the galaxy sky distribution. The central panels present the scatter plots of the sky localization area versus the error on $d_L$ (with a color scale giving information on the S/N). Finally, the bottom panels show the redshift and mass distribution of the detected GW events, as well as the distribution of the number of galaxies found within their localization volumes.
        
        The \OFour\ events are at redshift $z\lesssim 0.9$ with sky areas between a few and a few hundred square degrees. This typically results in more than $\sim 500$ potential host galaxies for 90\% of the events (Fig.~\ref{fig:sampleGW}, bottom-right panel). A particularly lucky event is present with a small localization area ($\sim2~\mathrm{deg^2}$) and high S/N ($\sim55$, see the mid-left panel of Fig.~\ref{fig:sampleGW}). While this event represents an outlier of the distribution that can be ascribed to a statistical fluctuation, it still represents a possibility with this configuration. For the \OFive\ events, the redshift distribution remains limited to $z\lesssim 1$ as a consequence of the higher S/N cut, while the larger and more sensitive detector network substantially improves the localization capabilities. This typically results in more than $\sim 50$ potential host galaxies for 90\% of the events (Fig.~\ref{fig:sampleGW}, bottom-right panel). In this configuration, there is a significant tail of events with just a few tens of galaxies, corresponding to sky localization regions of less than $1~\mathrm{deg^2}$ with an S/N that can exceed 100. Overall, while the number of galaxies in the localization volume depends on the assumption on the galaxy catalog employed, it is interesting to observe the $10\times$ reduction in the number of potential host galaxies between the O4 and O5 networks. This improvement, following from the much smaller localization volumes, is a key factor in obtaining improved dark siren constraints as discussed in Section~\ref{sec:results}.

        Ultimately, for the computation of the selection bias, we generate injection sets for both the \OFour\ and the \OFive\ scenarios with \texttt{GWFAST}, adopting the same selection cuts as for the GW catalogs. 
        The injections cover the same sky area as the catalogs and a distance range that arrives up to the detector horizon for the given $\mathrm{S/N}$ cuts.
        The injection set is made of $N_{\rm inj} = 2\times10^7$ and $4\times10^7$ events, resulting in about $1.5\times10^6$ and  $1\times10^6$ detected events in the \OFour\ and \OFive\ scenario, respectively. These are then used to estimate the selection bias as in Eq.~\ref{eq:bias}.

    \begin{table*}[t]
         \caption{Median and 68\% Confidence Level Interval of the Hyperparameters Resulting from Spectral-Only and Full Analyses for the \textit{O4-} and \textit{O5-like} Configurations and for Spectroscopic and Photometric Errors on the Galaxy Redshifts.} \label{tab:results_summary}
         \noindent\makebox[\textwidth]{%
         \begin{tabular}{{X{1.0cm}|X{2.4cm}X{2.4cm}X{2.4cm}|X{2.4cm}X{2.4cm}X{2.4cm}}}
             \hline
             \hline
     	 \multirow{2}{*}{} & \multicolumn{3}{c|}{\OFour\ (100 Events with S/N $>$ 12)} & \multicolumn{3}{c}{\OFive\ (100 Events with S/N $>$ 25)} \\  [2pt]
             \cline{2-7}
     		   & \Spectral\ & \Fullzp\ & \Fullzs\ & \Spectral\ & \Fullzp\ & \Fullzs\ \\ [2pt]
    			\hline
    		$H_0$ & $64^{+32}_{-23}\,(43\,\%)$ & $76^{+16}_{-12}\,(18\,\%)$ & $75.3^{+5.2}_{-4.9}\,(7\,\%)$ & $55^{+20}_{-16}\,(32\,\%)$ & $73^{+7}_{-6}\,(9\,\%)$ & $70.2\pm 0.8\,(1\,\%)$ \\  [2pt]
      		$\gamma$ & $1.6^{+1.3}_{-1.0}\,(67\,\%)$ & $1.8^{+1.1}_{-0.9}\,(54\,\%)$ & $1.8^{+1.4}_{-0.9}\,(62\,\%)$ & $2.2^{+1.9}_{-1.1}\,(68\,\%)$ & $2.3^{+1.6}_{-0.9}\,(56\,\%)$ & $2.3^{+1.3}_{-1.0}\,(50\,\%)$  \\  [2pt]
    		$\alpha$ & $3.6 \pm 0.4\,(10\,\%)$ & $3.7\pm 0.4\,(10\,\%)$ & $3.7 \pm 0.3\,(9\,\%)$ & $3.2\pm 0.3\,(8\,\%)$ & $3.3 \pm 0.3\,(8\,\%)$ & $3.3 \pm 0.3\,(8\,\%)$ \\ [2pt]
    		$\beta$ & $2.5^{+0.8}_{-0.7}\,(29\,\%)$ & $2.4^{+0.9}_{-0.7}\,(32\,\%)$ & $2.6\pm 0.7\,(29\,\%)$ & $1.9^{+0.6}_{-0.5}\,(28\,\%)$ & $2.0 \pm 0.6\,(29\,\%)$ & $1.8 \pm 0.6\,(30\,\%)$ \\ [2pt]
    		$\delta_m$ & $7.1^{+1.9}_{-3.2}\,(34\,\%)$ & $7.0^{+2.1}_{-3.1}\,(37\,\%)$ & $7.3^{+1.8}_{-2.5}\,(29\,\%)$ & $3.1^{+2.5}_{-1.8}\,(69\,\%)$ & $2.8^{+2.3}_{-1.6}\,(70\,\%)$ & $3.5^{+2.3}_{-1.7}\,(57\,\%)$ \\ [2pt]
    		$m_{\rm low}$ & $4.4^{+0.8}_{-0.7}\,(16\,\%)$ & $4.3^{+0.7}_{-0.8}\,(16\,\%)$ & $4.3^{+0.6}_{-0.5}\,(13\,\%)$ & $6.0^{+0.7}_{-1.1}\,(15\,\%)$ & $5.8^{+0.7}_{-1.1}\,(15\,\%)$ & $5.6^{+0.7}_{-1.0}\,(15\,\%)$ \\ [2pt]
    		$m_{\rm high}$ & $102^{+52}_{-21}\,(35\,\%)$ & $95^{+47}_{-17}\,(33\,\%)$ & $94^{+59}_{-15}\,(38\,\%)$ & $102^{+23}_{-13}\,(18\,\%)$ & $91.5^{+27.8}_{-7.2}\,(19\,\%)$ & $93.2^{+35.3}_{-7.1}\,(23\,\%)$ \\ [2pt]
    		$\mu_{\rm g}$ & $34.2^{+3.5}_{-4.8}\,(12\,\%)$ & $32.9^{+2.4}_{-3.5}\,(9\,\%)$ & $32.8^{+1.7}_{-2.7}\,(7\,\%)$ & $36.5\pm 2.7\,(7\,\%)$ & $34.1^{+1.3}_{-1.4}\,(4\,\%)$ & $34.40^{+0.99}_{-1.05}\,(3\,\%)$ \\ [2pt]
    		$\sigma_{\rm g}$ & $5.3^{+2.6}_{-1.8}\,(42\,\%)$ & $5.0^{+2.0}_{-1.6}\,(35\,\%)$ & $4.9^{+2.2}_{-1.4}\,(38\,\%)$ & $4.47^{+1.21}_{-0.84}\,(23\,\%)$ & $4.09^{+0.90}_{-0.65}\,(19\,\%)$ & $4.24^{+0.82}_{-0.71}\,(18\,\%)$ \\  [2pt]
                $\lambda_{\rm g}$ & $0.03^{+0.03}_{-0.01}\,(74\,\%)$ & $0.03^{+0.03}_{-0.01}\,(82\,\%)$ & $0.03^{+0.02}_{-0.01}\,(62\,\%)$ & $0.07^{+0.04}_{-0.03}\,(50\,\%)$ & $0.07^{+0.04}_{-0.03}\,(46\,\%)$ & $0.07^{+0.04}_{-0.03}\,(49\,\%)$ \\ [2pt]
            \hline
         \end{tabular}
         }
         \tablecomments{The relative uncertainties are given in parentheses. The rate parameters $k$ and $z_p$ are not included as they remain practically unconstrained.}
    \end{table*}

    \begin{figure*}[t]
        \centering
        \includegraphics[width=0.99\textwidth]{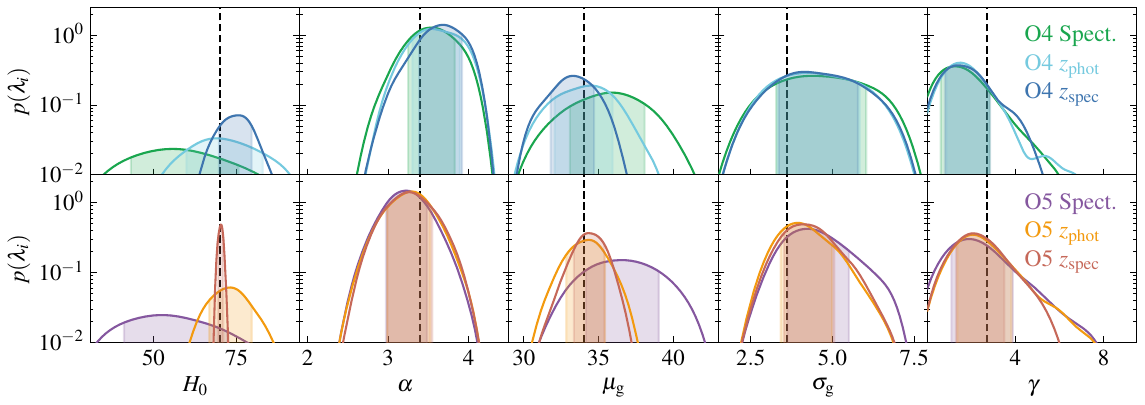}
        \caption{Marginalized posterior distributions of a selection of hyperparameters in the \OFour\ (\textit{upper panels}) and \OFive\ (\textit{lower panels}) detector networks. The analysis is performed with \CHIMERA{} in three setups, namely, spectral-only (Spect.) and spectral with the inclusion of a photometric ($z_{\rm phot}$) or spectroscopic ($z_{\rm spec}$) galaxy catalog.}
        \label{fig:1d-comparison}
    \end{figure*}

\section{Results}\label{sec:results}

    In this section, we report the results of the analyses of the \textit{O4-} and \textit{O5-like} configurations. For both detector networks, we consider three distinct analysis setups:
    \begin{itemize}
        \item \Fullzs: Full analysis using the parent catalog of 1.6 million galaxies derived from MICEv2 adopting spectroscopic redshift uncertainties as in Eq.~\ref{eq:zerr:def}. 
        \item \Fullzp: Full analysis as above, adopting broader (photometric) redshift uncertainties as in Eq.~\ref{eq:zerr:def}. 
        \item Spectral: No galaxy catalog information is incorporated and instead galaxies are assumed to be uniformly distributed in comoving volume. The constraints are based only on the assumption of the source-frame mass distribution.
    \end{itemize}
    In this way, we obtain a total of six configurations. We adopt wide uniform priors for all the hyperparameters (Table~\ref{tab:parameters}).
    The posterior distribution is sampled with affine-invariant Markov Chain Monte Carlo sampler \texttt{emcee} \citep{ForemanMackey2013} and the convergence is assessed ensuring that the number of samples is at least 50 larger than the integrated autocorrelation time for all the parameters. Results are reported using the median statistic with symmetric 68.3\% credible levels.

    The final constraints on all individual parameters are reported in Table~\ref{tab:results_summary}, while the marginalized posterior distribution for a selection of cosmological ($H_0$), mass ($\alpha$, $\mu_{\rm g}$, $\sigma_{\rm g}$), and rate ($\gamma$) parameters is shown in Fig.~\ref{fig:1d-comparison}. 
    In Fig.~\ref{fig:H0marginals} we show the relative uncertainty on $H_0$ in all configurations.
    Finally, in Appendix~\ref{app:full_corner} we show an example of the full corner plot. For all the six configurations, with \CHIMERA{} we recover the fiducial values with a typical deviation of $0.2~\sigma$, with fluctuations that can be ascribed to the specific realizations of the data sets.

    \subsection{Full O4- and O5-like scenarios}\label{sec:res:z_spec}
    
\begin{figure*}[t]
            \centering
            \includegraphics[width=0.7\textwidth]{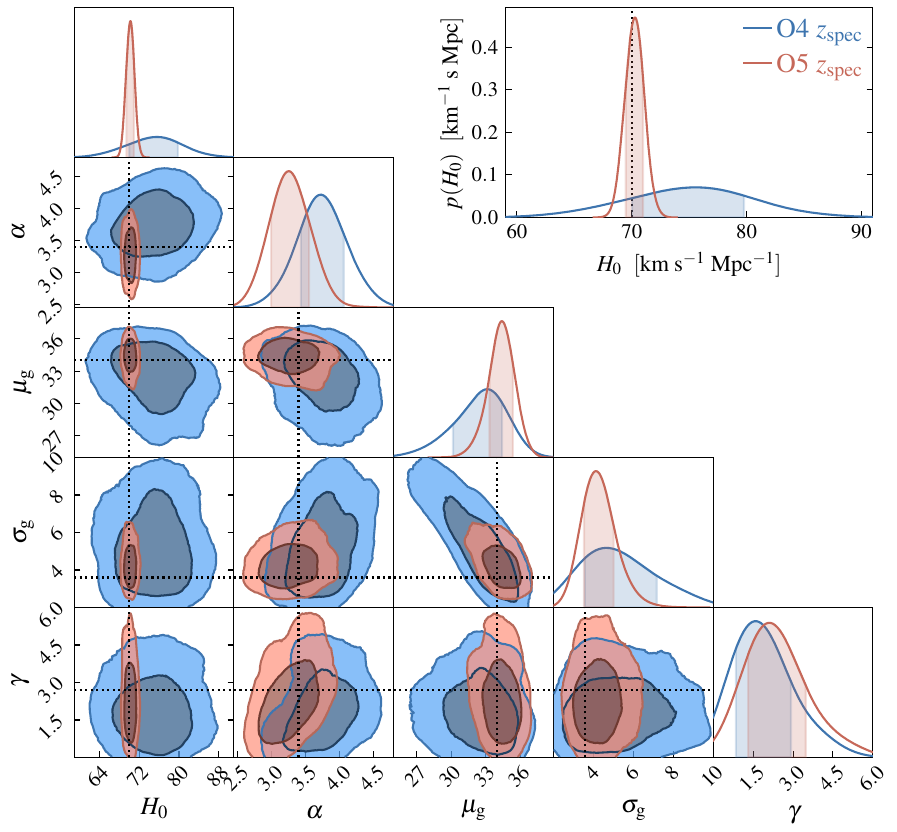}
            \caption{Joint cosmological and astrophysical constraints from the full standard sirens analysis of 100 BBHs in the \OFour\ (blue) and \OFive\ (red) configurations. Here we show, for illustrative purposes, the most relevant parameters, namely $\vlambda_{\rm c}=\{H_0\}$, the BBH mass function parameters $\vlambda_{\rm m}=\{\mu_{\rm g}, \sigma_{\rm g}\}$, and the rate evolution parameter $\vlambda_{\rm z}=\{\gamma\}$, while in App.~\ref{app:full_corner} we show for completeness the entire distribution of parameters. The contours represent the $1$ and $2\,\sigma$ confidence levels. The dotted lines indicate the fiducial values adopted.}
            \label{fig:triangle-O4_vs_O5-comparison}
        \end{figure*}

        We start by discussing the results for the best-case scenario, consisting of a complete galaxy catalog with spectroscopic redshift measurements. The marginal posterior distributions for the selection of parameters $ \{H_0, \alpha, \mu_{\rm g}, \sigma_{\rm g},\gamma \}$ are shown in Fig.~\ref{fig:triangle-O4_vs_O5-comparison}.
        
        We find that in 1~yr of observations in the \OFour\ configuration, the LVK data are able to constrain $H_0$ with $7\%$ uncertainty (at the $1\sigma$ level) from BBH in a combined cosmological and astrophysical population inference. This is a remarkable improvement with respect to the current constraints, as the analysis of the 42 BBHs at $\mathrm{S/N}>12$ observed so far yields a $46\%$ measurement of $H_0$ \citep{Mastrogiovanni2023}. To arbitrate the Hubble tension, percent-level measurements are required. If we assume that the uncertainty on $H_0$ scales as $1/\sqrt{N}$, in the O4 configuration it is not possible to reach $1\%$ in the planned schedule of about 2~yr.

        In contrast, with the \OFive\ configuration it is possible to reach $1\%$ uncertainty in about 1~yr of operation. We stress that this is a best-case scenario, relying on having a complete catalog of potential hosts.
        In general, the actual completeness can vary among different galaxy surveys and galaxy types. For example, with Euclid it would change between the photometric or the spectroscopic survey mode and the north or south direction, requiring a more detailed assessment in a future study. 
        Moreover, even if we marginalize on the astrophysical parameters, the results still rely on an assumption of the functional form of the BBH mass distribution. This dependence should be investigated in more detail in the future.
        On the other hand, this analysis is based only on the BBH population, and further improvements can be obtained including NSBH and BNS events and their potential EM counterparts.

\begin{figure}[b]
            \vspace{1em}
            \includegraphics[width=0.48\textwidth]{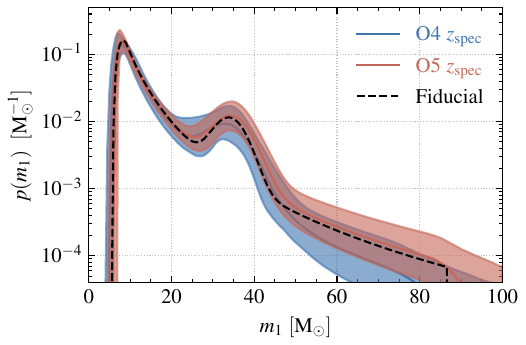}
            \caption{Posterior on the Power Law + Peak BBH primary mass function from the full standard analysis of 100 BBH in the \OFour\ (blue) and \OFive\ (red) configurations. The colored lines and bands represent the median and 90\% credible level. The black dashed line is the mass function evaluated at the fiducial hyperparameters.}
            \label{fig:MF_reconstruction}
        \end{figure}

        We now move to the population hyperparameters. Currently, in population studies the cosmology is typically fixed \citep[e.g.][]{Abbott2023Population}. Here we study how well the fiducial models are recovered when taking into account potential correlations with cosmological hyperparameters. In Fig.~\ref{fig:MF_reconstruction} we show the reconstructed primary mass distribution. For both O4 and O5 scenarios, the Gaussian peak at $34~\msun$ is clearly visible and its mean value $\mu_\mathrm{g}$ is recovered with a precision of $7\%$ and $3\%$, respectively. The second best-constrained mass parameter is the slope $\alpha$ of the primary BBH mass distribution that is recovered with fractional uncertainties of $15\%$ and $13\%$, respectively. Overall, for the mass function parameters these are small improvements with respect to the full cosmological and astrophysical inference with the current GWTC-3 catalog, which yields $\sigma_{\mu_\mathrm{g}}/\mu_\mathrm{g}\sim 13\%$ and $\sigma_\alpha/\alpha \sim 11\%$ \citep[see][]{Mastrogiovanni2023}. The same is true for the population analysis at fixed cosmology, which gives $\sigma_{\mu_\mathrm{g}}/\mu_\mathrm{g}\sim 9\%$ and $\sigma_\alpha/\alpha \sim 11\%$ \citep{Abbott2023Population}. 
        Similarly, the constraints on the rate parameter $\gamma$ remain essentially unchanged with respect to the current knowledge, even when considering the transition from O4 to O5. This can be explained by the higher S/N threshold adopted in our O5 catalog, resulting in the same number of GW events that map a redshift range comparable to that of O4. 
        
        In conclusion, we recall that our results are based on the full astrophysical and cosmological analysis of the best 100 GW events detectable in 1~yr for each configuration. Population studies typically benefit from the inclusion of all confident events detected and are carried out by fixing the cosmological parameters \citep[e.g.,][]{Abbott2023Population}. In this sense, our population constraints should not be taken as representative of the overall performance of O4 and O5.

        \begin{figure}[t]
            \centering
            \includegraphics[width=0.48\textwidth]{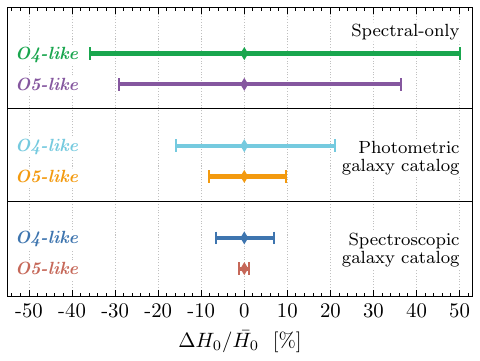}
            \caption{Relative uncertainty on $H_0$ obtained from spectral-only and full standard sirens analysis of 100 BBHs in the \textit{O4-} and \textit{O5-like} network configurations, including a complete galaxy catalog with photometric or spectroscopic redshift uncertainties.}        
            \label{fig:H0marginals}
        \end{figure}
    
    \subsection{Spectroscopic vs. photometric galaxy catalog}\label{sec:res:z_phot}

        While obtaining a complete spectroscopic catalog poses challenges and awaits future facilities, ongoing surveys such as DES and Euclid are already building extensive photometric galaxy catalogs. The full \zp\ and \zs\ configurations for both the \OFour\ and \OFive\ configurations are compared in Fig.~\ref{fig:1d-comparison}.
        
        With photometric redshifts, the constraint on $H_0$ for O4 is notably less accurate, with a measurement uncertainty that is three times greater ($\sigma_{H_0}/H_0 \approx 18\%$) compared to the spectroscopic approach. In the case of O5 this factor increases to 9 ($\sigma_{H_0}/H_0\approx 9\%$). Interestingly, this shows that from 100 O5 events at $\mathrm{S/N}>25$ it is not possible to achieve percent-level precision on $H_0$ using a photometric catalog, even under the assumption of completeness. 
        Of course, one may consider lowering the S/N threshold to include more events; however, such events would also have much larger localization volumes, and thus a large number of potential hosts. This would likely limit their additional constraining power. Some information can be still retrieved for the mass distribution, whose reconstruction benefits from a larger sample; however, in the next section we will show that the constraints obtained in the absence of a galaxy catalog with our sample are only marginally worse than what can be obtained from a sample with a much lower S/N cut. This seems to suggest that it would be difficult to reach percent-level accuracy.

        Another interesting result concerns the comparison of the \Fullzp\ \OFive\ and \Fullzs\ \OFour\ configurations (see Fig.~\ref{fig:H0marginals}). We find that considering a galaxy catalog with spectroscopic redshift uncertainties in the \OFour\ scenario, we are able to achieve a more precise constraint on $H_0$ compared to having a larger LVK detector network at O5 design sensitivity with a photometric galaxy catalog. This occurs despite the factor 10 improvement in the GW localization volume (see Fig.~\ref{fig:sampleGW}). Overall, these findings underline the importance of mapping the GW localization volume - at least for well-localized events - with dedicated spectroscopic surveys.

    \subsection{Spectral-only analysis}\label{sec:res:spectral}

        In the absence of the galaxy catalog information (spectral sirens case), the cosmological constraints are determined by the capability of the source-frame mass function to break the mass-redshift degeneracy. 
        
        In our spectral siren analysis, the Hubble parameter is recovered at $\sigma_{H_0} / H_0 \approx 43\%$ in \OFour\ and $\sigma_{H_0} / H_0 \approx 32\%$ in \OFive\ configuration. Our results are in good agreement with those of \cite{Mancarella2022} and \cite{Leyde2022}. In particular, \cite{Leyde2022} find $38\%$ ($24\%$) uncertainty on $H_0$ using $\mathrm{S/N}>12$ spectral siren events for O4 (O5) obtaining a total of 87 (423) events. When comparing with our results for O5, we must consider that we applied a higher $\mathrm{S/N}>25$ threshold, resulting in a factor of $4$ difference in the number of detected events. It is interesting to note that this large difference results only in a marginal improvement in the constraining power, suggesting that the constraints are mainly driven by the events with higher S/N.

        In general, we conclude that the analysis of 100 well-localized GW events with a complete galaxy catalog with spectroscopic measurements would provide better constraints with respect to a pure spectral siren analysis under the assumed mass functions. This result also holds true in comparison to the 5~yr results of O5 spectral sirens by \cite{Farr2019}, providing a relative error on $H(z)$ of $\approx 3\%$ at a pivot redshift of $z\simeq0.8$. We note however that the authors assumed a much more optimistic event rate and a mass function with sharper edges, in line with the knowledge of the BBH population at the time.

    \subsection{The \texorpdfstring{$H_0-\mu_{\rm g}$}{H0-mu} correlation}\label{sec:res:correlations}

\begin{figure}[t]
            \centering
            \includegraphics[width=\hsize]{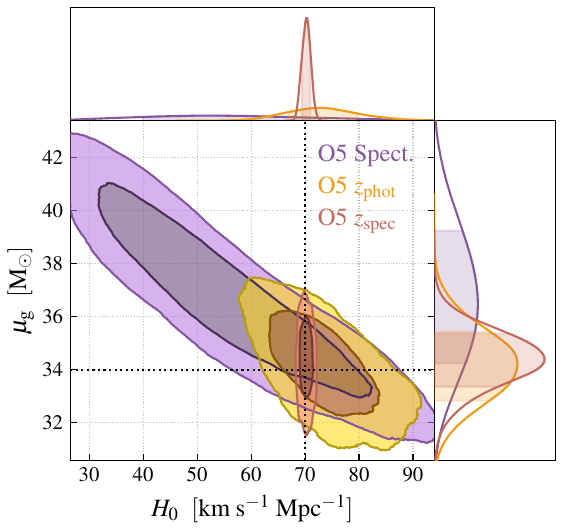}
            \caption{Constraints on the $H_0-\mu_{\rm g}$ plane for the \OFive\ scenario in the case of spectral, \Fullzp, \Fullzs\ analyses with \CHIMERA{}.}
            \label{fig:H0mug}
        \end{figure}

        In the absence of a galaxy catalog, there is a strong correlation between $H_0$ and $\mu_{\rm g}$, which drives the constraint on $H_0$. In this section, we comment on this correlation and compare the three scenarios considered in the paper. Figure~\ref{fig:H0mug} shows the constraints in the $H_0-\mu_{\rm g}$ plane in the \OFive\ configuration.
        
        A relatively strong anticorrelation is present between $\mu_{\rm g}$ and $H_0$ in both the Spectral and \Fullzp\ analyses. In terms of the Spearman's rank correlation coefficient computed from the MCMC chains, these correspond to $\rho \sim -0.8$ and $\rho \sim -0.6$, respectively.
        
        This is a consequence of the fact that, in the absence of a precise redshift measurement, a higher value of $H_0$ would move the events with a given luminosity distance at higher $z$, requiring smaller values of the source-frame masses to reproduce the observed data (see Eq.~\ref{eq:msourcedet}). This leads to a smaller value of $\mu_{\rm g}$. This anticorrelation is observed also in the latest GWTC-3 analyses \citep{Abbott2023}. Having a complete spectroscopic galaxy catalog allows instead to break the $H_0-\mu_{\rm g}$ degeneracy, constraining $H_0$ with higher precision. This is visible in Fig.~\ref{fig:H0mug} where no significant correlation is present in the \Fullzs\ case. Interestingly, in the case where only photometric redshifts are available, this result shows that the information coming from the mass distribution still plays a role in the measurement of $H_0$ and has therefore to be accurately modeled. In particular, Fig.~\ref{fig:H0mug} shows that a wrong assumption on the value of the mass scale would lead to a bias on $H_0$ even in the presence of a complete photometric galaxy survey.

\section{Conclusions}\label{sec:conclusions}

    In this work, we study the scientific potential of dark sirens in the context of a joint astrophysical and cosmological parameter inference including galaxy catalogs. Here we summarize the main results of the paper.
    
    \begin{enumerate}
        \item We present \CHIMERA{}\footnote{Available at \url{https://github.com/CosmoStatGW/CHIMERA}}, a novel Python code for GW cosmology combining information from the population properties of compact binary mergers and galaxy catalogs. The code is designed to be accurate in different scenarios, from the spectral siren (population inference), to the dark siren (galaxy catalog) and the bright siren (redshift precisely measured after the identification of an EM counterpart) methods, and efficient for large galaxy catalogs.
    
        \item We generate two catalogs of BBH events for detector networks representative of the LIGO-Virgo-KAGRA O4 and O5 observing runs. We start from a parent catalog of galaxies extracted from the MICEv2 mock galaxy catalog, which we select with a cut in stellar mass $\log M_\star/\msun > 10.5$. From this catalog, we generate BBH events with the \texttt{GWFAST} code, considering the 100 best events that can be detected over approximately 1~yr of observation, namely, $\mathrm{S/N}>12$ BBHs for the \OFour\ configuration and 100 $\mathrm{S/N}>25$ BBHs for the \OFive\ configuration. We then associate the galaxy redshift catalog with two different uncertainties on the redshift, representative of the case of having a catalog of photometric and spectroscopic redshifts.
        
        \item We find that in the best-case scenario of a complete spectroscopic galaxy catalog, the Hubble constant $H_0$ can be constrained to 7\% with \OFour\ and 1\% with \OFive\ in about a year of observations, demonstrating the strong potential of joint standard sirens and galaxy catalog analyses in the context of the Hubble tension using novel independent probes.
    
        \item With a photometric galaxy catalog, the constraints on $H_0$ are notably weaker, i.e. by a factor of $\sim 3$ for \OFour\ and $\sim 9$ for \OFive. In this case, the correlation between $H_0$ and the mass scales in the BBH population - which is intrinsic in the absence of a galaxy catalog - is not fully broken. It is therefore crucial to accurately model the mass distribution to avoid biases in the cosmological parameters.

        \item Interestingly, we find that the \OFour\ configuration with a spectroscopic catalog provides a more precise measurement of $H_0$ than the \OFive\ configuration with photometric redshifts. This points to the importance of having spectroscopic redshift measurements for GW cosmology, in the absence of which the potential of a factor $\sim10$ improvement in the localization capabilities between \textit{O4-} and \OFive\ GW detector networks could be completely lost.
    \end{enumerate}

    The results obtained in this work are based on some simplifying assumptions. First of all, the catalog of potential hosts that we employ includes only the more massive galaxies, in line with the idea that the binary merger rate is traced by stellar mass. Second, the catalog is assumed to be complete. Despite these limitations, the crucial role that the spectroscopic surveys of galaxies can have in GW cosmology is already evident. Moreover, while obtaining a complete catalog of less massive galaxies is challenging at present, this will become a concrete possibility in the coming years thanks to ongoing (e.g., DESI, \citealt{desi}; and \textit{Euclid}, \citealt{Laureijs2011}), future (e.g., WFIRST, \citealt{Akeson2019}; LSST, \citealt{Collaboration2009}) and proposed (e.g., WST, \citealt{WST}) large galaxy surveys. In this context, more research is needed to further study the impact of incompleteness in the observed catalog. 
    
    Potential improvements may come from two directions. Spectroscopic targeted searches may be carried out, also a posteriori, in the localization volumes of best-localized events. To study the feasibility of this approach, a comprehensive study should be carried out, considering both the observational requirements and the optimal GW localization volumes within which this could be done. Even if this approach is not feasible, the additional information from the observational and physical galaxy properties in combination with binary population studies \citep[e.g.][]{Santoliquido2022, Vijaykumar2023a} can be included in the inference, potentially improving the constraints. 

    From the GW data side, a second assumption, which mostly impacts the photometric and spectral-only analyses, is the specific distributions for the binary population. In fact, even if we marginalize on the astrophysical parameters, the results still rely on the functional form of the BBH mass distribution and the presence of features that can be used as standard rulers to obtain cosmological results. This dependence should be investigated in more detail in the future. We also emphasize that the GW catalogs analyzed in this work are built assuming a 100\% duty cycle, although, in reality, detectors may undergo temporary maintenance and upgrade phases, requiring modeling of the specific duty cycle for each detector. Moreover, we recall that these results are obtained assuming at least three GW detectors working at nominal sensitivity, which is crucial in order to achieve a sufficiently precise localization. 
    
    In the future, it will be interesting to expand this analysis to 3G networks, such as the Einstein Telescope \citep{Punturo2010, Branchesi2023}, LISA \citep{AmaroSeoane2017}, and Cosmic Explorer \citep{Reitze2019}, exploring deeper galaxy catalogs needed to properly cover the detectability range of those detectors. We will address the case of future GW experiments in a separate study.
    
    To conclude, our results highlight that the synergy between future GW detectors and large galaxy surveys is crucial to fully attain the potential of dark sirens analyses. In particular, they suggest the importance of targeted spectroscopic campaigns to map the galaxy distribution within the GW localization volumes, at least for the best-localized events.

\begin{acknowledgments}
    We thank Simone Mastrogiovanni, Sofia Contarini, Massimiliano Romanello, Giorgio Lesci, Alberto Traina, and Laura Leuzzi for useful discussions. We acknowledge the use of computational resources from the parallel computing cluster of the Open Physics Hub (\url{https://site.unibo.it/openphysicshub/en}) at the Department of Physics and Astronomy in Bologna. N.B. and M.Mor. acknowledge support from MIUR, PRIN 2017 (grant 20179ZF5KS). M.Mor. and A.C. acknowledge the grants ASI n.I/023/12/0 and ASI n.2018-23-HH.0. A.C. acknowledges the support from grant PRIN MIUR 2017 - 20173ML3WW\_001. M.Mor. acknowledges the support from grant PRIN MIUR 2022 (grant 2022NY2ZRS\_001)
    M.T. acknowledges the funding by the European Union -  NextGenerationEU, in the framework of the HPC project - ``National Centre for HPC, Big Data and Quantum Computing'' (PNRR - M4C2 - I1.4 - CN00000013 - CUP J33C22001170001). 
    The work of M.Manc. is supported by European Union's H2020 ERC Starting Grant No.~945155--GWmining, Cariplo Foundation Grant No.~2021-0555, the ICSC National Research Centre funded by NextGenerationEU, and MIUR PRIN Grant No. 2022-Z9X4XS. The work of F.I. and M.Mag. is supported by the Swiss National Science Foundation, grant 200020$\_$191957, and by the SwissMap National Center for Competence in Research. The views and opinions expressed are those of the authors only and do not necessarily reflect those of the European Union or the European Commission. Neither the European Union nor the European Commission can be held responsible for them.
    This work is (partially) supported by ICSC - Centro Nazionale di Ricerca in High Performance Computing, Big Data and Quantum Computing, funded by European Union - NextGenerationEU.
\end{acknowledgments}

\software{\texttt{GWFAST} \citep{Iacovelli2022-GWFAST}, \texttt{MGCosmoPop} \citep{Mancarella2022}, \texttt{ChainConsumer} \citep{Hinton2016}, \texttt{NumPy} \citep{Harris2020}, \texttt{matplotlib} \citep{Hunter2007}.}

\appendix

\section{\CHIMERA{}: code implementation}\label{app:code}

\begin{figure*}
        \centering
        \includegraphics[width=0.8\textwidth]{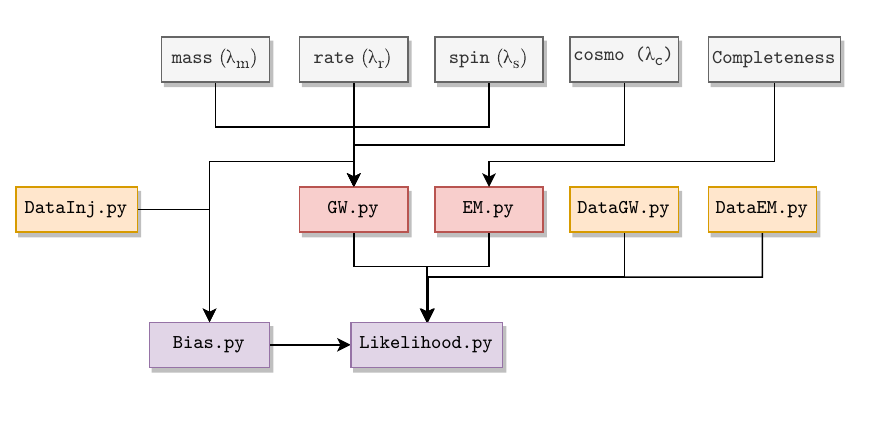}
        \vspace*{-2em}
        \caption{Workflow of \CHIMERA{}. The main modules include functions related to the full likelihood computation (violet), computation of probabilities (red), data file I/O (yellow), and model functions (gray). }
        \label{fig:CHIMERA-structure}
    \end{figure*}

    \href{https://github.com/CosmoStatGW/CHIMERA}{\CHIMERA{} \faGithub} is a new, flexible Python code that enables jointly fitting the cosmological and astrophysical population parameters by using information from galaxy catalogs. The code is designed to be accurate for different scenarios, encompassing bright, dark, and spectral siren methods, and computationally efficient in view of next-generation GW observatories and upcoming galaxy surveys. The ultimate goal of \CHIMERA{} is to compute the result of Eq.~\ref{eq:like_full} in a computationally efficient way in view of next-generation GW observatories and upcoming galaxy surveys. For this reason, we use the LAX-backend implementation and Just-in-time computation capabilities of \texttt{JAX} \citep{JAX}. The main functions are developed starting from \texttt{DarkSirensStat} \citep{Finke2021_DarkSirens} and \texttt{MGCosmoPop} \citep{Mancarella2022}. The workflow of \CHIMERA{} is shown in Fig.~\ref{fig:CHIMERA-structure}. 

    In this appendix, we provide an overview of the code. As standard units for \CHIMERA{}, we adopt radians for the angles and gigaparsecs for the distances. Data sets are given in the form of dictionaries of \texttt{NumPy} arrays. All model functions accept one argument in the form of a dictionary containing the related parameters.
    
    \subsection{Full likelihood}
    
        The core modules of \CHIMERA{} are \textcolor[HTML]{9673a6}{\texttt{Likelihood.py}} and \textcolor[HTML]{9673a6}{\texttt{Bias.py}}. In particular, the file \textcolor[HTML]{9673a6}{\texttt{Likelihood.py}} contains classes to compute the product of the integrals in Eq.~\ref{eq:like_full}, while \textcolor[HTML]{9673a6}{\texttt{Bias.py}} allows for the computation of the selection bias term $\xi(\vlambda)$. The computation of the likelihood uses functions to analyze the GW (from \textcolor[HTML]{b85450}{\texttt{GW.py}}) and EM (from \textcolor[HTML]{b85450}{\texttt{EM.py}}) information. These modules contain specific methods to perform all the preliminary computations that do not change during the likelihood evaluation (a more extended discussion can be found in the next paragraph). Data are loaded with specific classes present in \textcolor[HTML]{d79b00}{\texttt{DataGW.py}} (e.g., \texttt{DataGWMock} and \texttt{DataLVK}) and \textcolor[HTML]{d79b00}{\texttt{DataEM.py}} (e.g., \texttt{MockGalaxiesMICEv2} and \texttt{GLADEPlus}), respectively. The algorithm to compute the \texttt{Likelihood} proceeds as follows:
        \begin{enumerate}
            \item First, the class \texttt{MockLike} stores all the population models (\texttt{model\_*}), GW data (\texttt{data\_GW*}), galaxy data (\texttt{data\_GAL*}), pixelization parameters (\texttt{nside\_list}, \texttt{npix\_event}, \texttt{sky\_conf}), and integration parameters (\texttt{z\_int\_H0\_prior}, \texttt{z\_int\_sigma}, \texttt{z\_int\_res}).
        
            \item The \texttt{GW} class is initialized and the pixelization and redshift grids are precomputed. To optimize the computation in the case of large galaxy catalog analysis, the code not only restricts the sky localization, but also the redshift integration grid. The first task is done starting from posterior distributions in RA and Dec. The user can choose the approximate number of pixels desired for each event (\texttt{npix\_event}) to be found within a confidence level ellipse (\texttt{sky\_conf}), given a list of possible pixelizations (\texttt{nside\_list}). \CHIMERA{} optimizes the pixelization of each event (in particular, the \textit{nside} parameter of HealPix) to obtain the number of pixels closest to \texttt{npix\_event}. The second task must take into account that by varying the cosmology, also $\mathcal{K}_\mathrm{gw}$ varies, so to avoid biases, the integration grid must encompass all the redshift ranges explored during the inference. This is obtained starting from the GW posteriors on $d_L$ and defining the range inside \texttt{z\_int\_sigma} standard deviations at a resolution of \texttt{z\_int\_res} spanning all the range\footnote{In practice, the grid starts at $z(d_L-N\,\sigma_{d_L}\given H_{0,\mathrm{min}})$ and ends at $z(d_L+N\,\sigma_{d_L}\given H_{0,\mathrm{max}})$, where $N$ is the number of standard deviations (\texttt{z\_int\_sigma}) and $[H_{0,\mathrm{min}},H_{0,\mathrm{max}}]$ the interval of $H_0$ spanned in the inference.} of $H_0$ explored.
        
            \item The \texttt{EM} class is initialized, and the quantity $p_{\rm cat}(z, \hat{\Omega} \given \vlambda_{\rm c})$ (Eqs.~\ref{eq:pcat_complete} and \ref{eq:pcat_zprior_onegal}) is precomputed pixel by pixel on the redshift grids. At this point, based on the included catalog (e.g., GLADE+, MICEv2), it is possible to activate catalog-related tasks (e.g., luminosity cut, or associate user-defined redshift uncertainties). Similarly,
            if defined, the completeness is computed, and the pixelized completeness function $P_{\rm compl}(z)$ (see the discussion in the above paragraph) is stored in the class to be accessible during the inference. 
        \end{enumerate}
        The pixelization approach and the precomputation of $p_\mathrm{gal}$ are essential for next-gen galaxy surveys. The first allows us to reduce the dimension, combining the probability of $10-10^3$ galaxies in one single pixel ensuring granularity in the sky analysis. The second allows us to avoid the computation of $p_\mathrm{gal}$ at each step of the MCMC inference. While this last approximation is not a problem for $H_0$ inference since both the numerator and denominator of Eq.~\ref{eq:pcat_zprior_onegal} have a $H_0^3$ dependence, the impact of a possible bias on $\Omega_m$ should be carefully assessed when future analyses using events at higher $z$ will be carried out. The algorithm to compute the selection Bias proceeds as follows:
        \begin{enumerate}
            \item First of all, the class \texttt{Bias} stores all the population models (\texttt{model\_*}), the directory of the GW injection catalog data (\texttt{file\_inj}), and the S/N threshold to be applied. Optional arguments also include the catalog interpolant function. If not given, the bias is evaluated for a uniform in comoving volume galaxy distribution (i.e., $\propto\mathrm{d}{V_C}/\mathrm{d}{z}$).
            \item The injection catalog is loaded by applying the chosen S/N cut. It is important to ensure that this cut is equivalent to the one adopted when creating the catalog of GW events to analyze.
        \end{enumerate}
        Thus, following Eq.~\ref{eq:hyperlike}, the full likelihood is conveniently derived by first calculating the log-likelihood for all the events and then subtracting the log-bias term, which is computed only once and multiplied by the number of events. The computation of the likelihood and selection bias is performed by calling the \texttt{.compute()} methods or, in logarithmic form, the \texttt{.compute\_ln()} methods. In the latter case, the models should also be given in logarithmic form.

    \subsection{Included models}
    
        Currently, \CHIMERA{} includes the following population models:
        \begin{itemize}
            \item Mass distributions (\texttt{mass.py}): \texttt{logpdf\_TPL} (Truncated Power Law), \texttt{logpdf\_BPL} (Broken Power Law), \texttt{logpdf\_PLP} (Power Law + Peak), \texttt{logpdf\_PL2P} (Power Law + 2 Peaks). 
            \item Rate evolutions (\texttt{rate.py}): \texttt{logphi\_PL} (Power Law), \texttt{logphi\_MD} (Madau-like).
            \item Spin distributions (\texttt{spin.py}): \texttt{logpdf\_G} (Gaussian), \texttt{logpdf\_U} (Uniform).
            \item Cosmological models (\texttt{cosmo.py}): fLCDM, fLCDM modified gravity
        \end{itemize}
        All the above functions accept parameters in the form of dictionaries, e.g. \texttt{lambda\_cosmo = \string{\codestring{H0}:70.0, \codestring{Om0}:0.3\string}}.

\section{Choices for the analysis of GW170817}\label{app:GW170817}

    In this appendix, we provide details about the analysis setup for the example of GW170817 in Section~\ref{sec:implementation}.
    
    We use the GLADE+ galaxy catalog \citep{Dalya2022}, which includes data from multiple galaxy surveys (GWGC, 2MPZ, Two Micron All Sky Survey, HyperLEDA, WISExSCOSPZ, and Sloan Digital Sky Survey DR16Q quasars) and is largely employed in standard sirens analyses. As done in previous analyses, we select galaxies with a luminosity above a certain threshold to obtain a more complete catalog of potential hosts. The implicit assumption is that the true host lies above the chosen threshold. This is well motivated by the fact that the luminosity is a reasonable proxy for the stellar mass (in K band) and the star formation rate (in B band), therefore more luminous galaxies have a higher chance to host mergers. 
    In this case, we use $K$-band luminosities with a threshold $L_K>0.02\,L_K^\ast$, where $L_K^\ast\simeq 1.1\times 10^{11}~\msun$ \citep{Kochanek2001} denotes the characteristic $K$-band luminosity. In this analysis we weight each galaxy by its luminosity, $w_g\propto L_k$ (see Eq.~\ref{eq:pcat_complete}). The completeness is computed with the mask method, as outlined in Section~\ref{sec:implementation}. The sky is pixelized with a size of 0.83 deg$^2$ and pixels are organized into nine distinct masks based on the galaxy counts in each pixel. For each mask, the completeness fraction $P_c(z, \hat\Omega)$ is computed by comparing the luminosity density to a reference Schechter function with parameters from \cite{Kochanek2001}. We adopt the following population model: flat mass distribution between $1$ and $3~\msun$ and power-law rate evolution with slope $\gamma=2.7$. Then, we fix the population parameters and study the posterior on $H_0$. For the inference, we consider each galaxy within the $90\%$ GW sky localization area (about 28 deg$^2$) as a potential host (Fig.~\ref{fig:GW170817}, left).

\section{Comparison of \CHIMERA{} with \texttt{MGCosmoPop}}\label{app:MGCP}
    
    We compare \CHIMERA{} in the spectral-only case (population analysis) with \texttt{MGCosmoPop}.\footnote{Available at \url{https://github.com/CosmoStatGW/MGCosmoPop}} The code is presented and used in \cite{Mancarella2022} to provide joint cosmological and astrophysical constraints, including deviation from general relativity, in the empty catalog case. The codes feature similar model function implementations, except for the cosmology class, which has been rewritten in \texttt{CHIMERA} to improve computational efficiency. We compare these functions across a wide range of parameters, finding machine-level differences. 
    
    On the contrary, the two codes feature distinct implementations of the likelihood evaluation. In \texttt{MGCosmoPop} the integral in Eq.~\ref{eq:hyperlike} is performed with a MC approach, while in \CHIMERA{} we pixelize the sky, evaluate the redshift prior inside each pixel, and perform the integral on the redshift grid (see Fig.~\ref{fig:CHIMERA_sketch}). While this approach has the advantage of improving the computational efficiency with large galaxy catalogs \citep{Gray2022} and the evaluation of their completeness \citep{Finke2021_DarkSirens}, it is crucial to perform a comparative analysis to better assess potential biases.

\begin{figure*}
        \centering
        \includegraphics[width=0.8\textwidth]{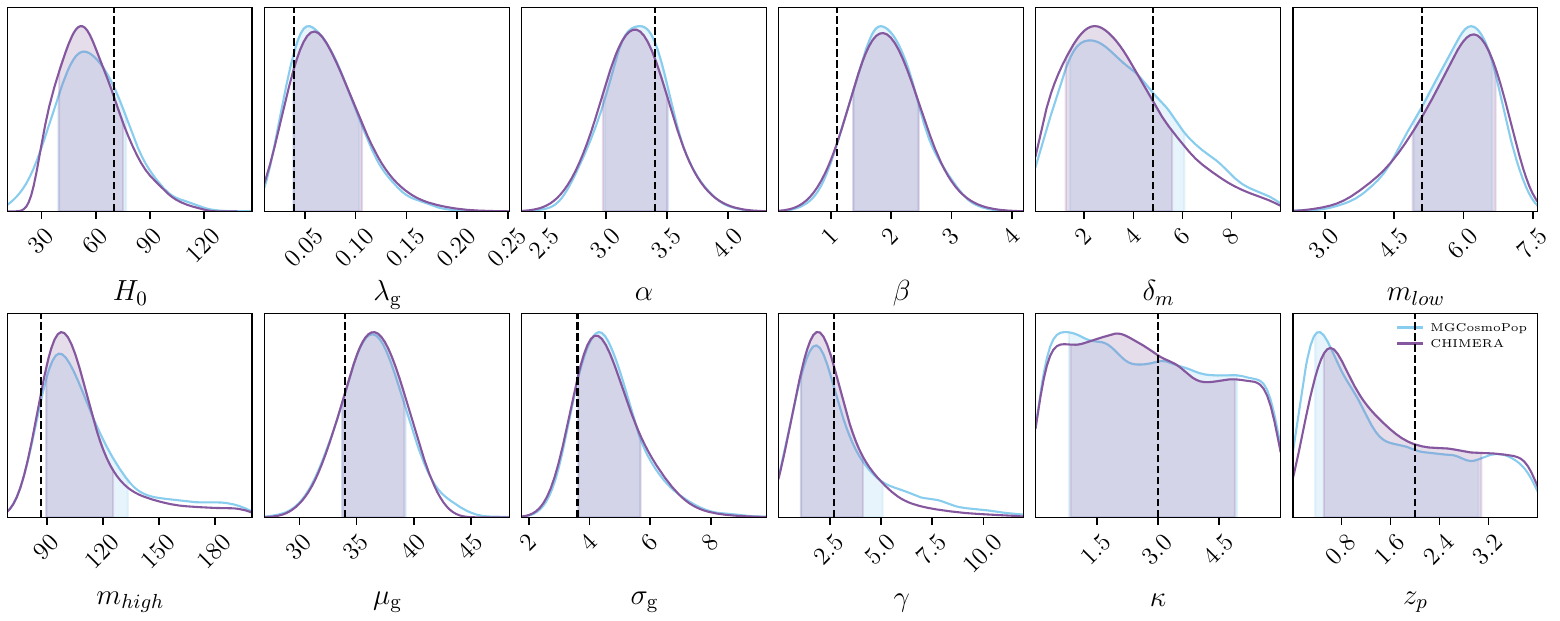}
        \caption{Posterior distributions obtained in the case of the spectral siren analysis with \CHIMERA{} and \texttt{MGCosmoPop} using the \OFive\ catalog.  }\label{fig:CHIMERAvsMGCP}
    \end{figure*}

    We use the set of 100 BBH events at $\mathrm{S/N}>25$ presented in Section~\ref{sec:GWsample}. We assume a flat $\Lambda$CDM cosmology, PLP mass distribution, and Madau-like rate evolution, with parameters and priors reported in Table~\ref{tab:parameters}. In \texttt{CHIMERA}, we set \texttt{npix\_event} to 15 and \texttt{nside} to $[2^n]$ with $n=(3,...,9)$ so that the pixelization of each event is automatically adjusted to have approximately 15 pixels in the 90\% credible sky area. Then, we perform full MCMC analyses with \texttt{CHIMERA} and \texttt{MGCosmoPop}. The results are shown in Fig.~\ref{fig:CHIMERAvsMGCP}. Overall, the posteriors obtained with the two codes are in excellent agreement.

\section{Full corner plot}\label{app:full_corner}
\begin{figure*}
        \centering
        \includegraphics[width=0.82\textwidth]{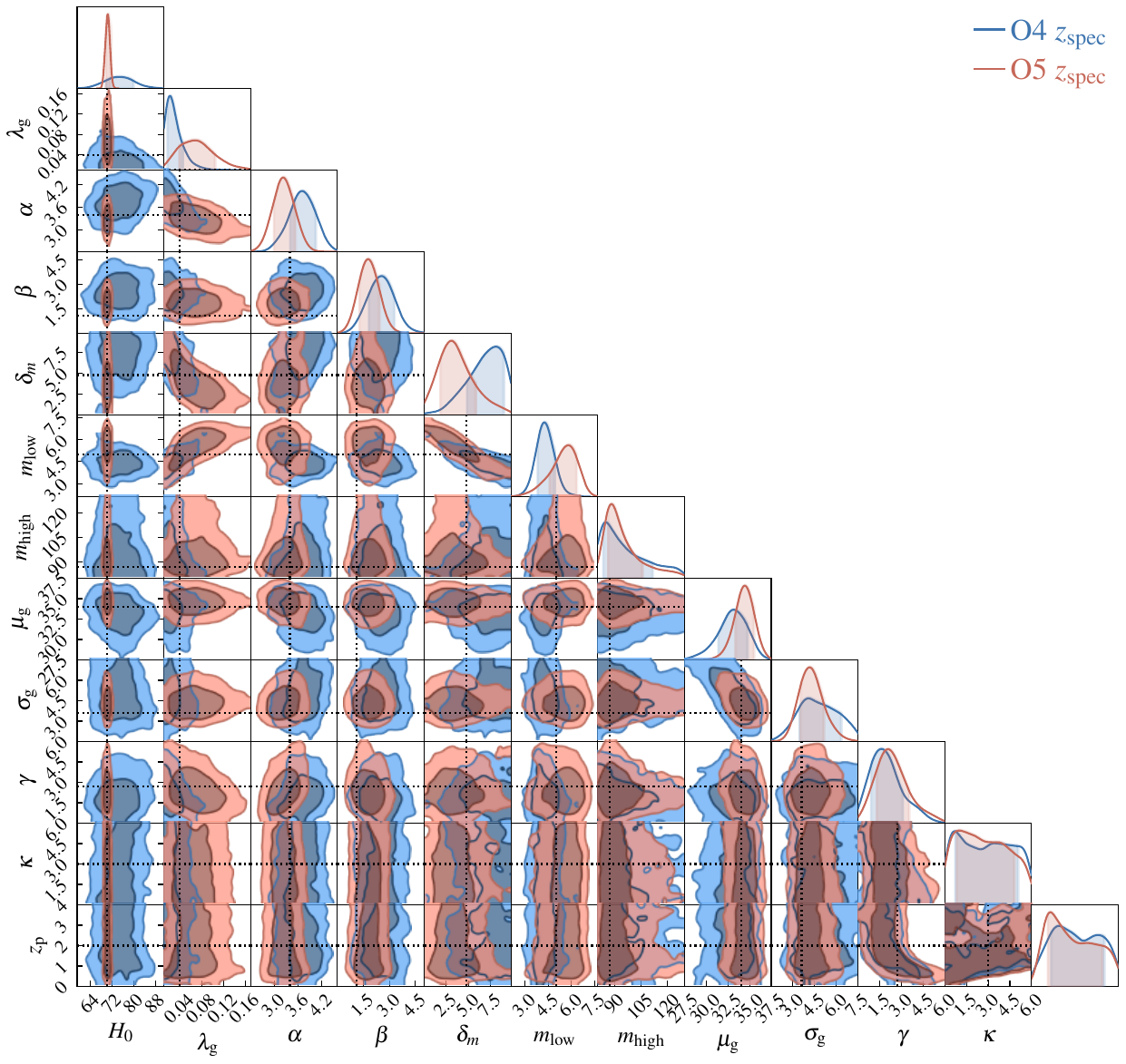}
        \caption{Joint cosmological and astrophysical constraints from the full standard sirens analysis of 100 BBHs in the O4 (blue) and O5 (red) configurations for the full parameter list described in Table~\ref{tab:parameters}. The contours represent the 68\% and 95\% confidence levels. The dotted lines indicate the fiducial values adopted.}
        \label{fig:mcmc_full}
    \end{figure*}

    In Fig.~\ref{fig:mcmc_full} we show the full corner plot with posteriors on the 12 parameters $\vlambda$ in the case of galaxy catalog with spectroscopic redshift measurements. In this scenario, as in the other four configurations (see Table~\ref{tab:results_summary}), we are able to recover the fiducial values for almost all of the hyperparameters within the 68\% credible levels.

\clearpage
\bibliography{main}{}
\bibliographystyle{aasjournal}

\end{document}